\documentclass[
reprint,
 amsmath,amssymb,
 aps,
showtitles
]{revtex4-2}

\usepackage{graphicx}
\usepackage{dcolumn}
\usepackage{bm}
\usepackage{hyperref}
\usepackage[mathlines]{lineno}
\usepackage{lipsum}  
\usepackage{epsfig, amssymb, amsmath, amsthm, braket, physics, xcolor}
\usepackage{cleveref}
\usepackage[normalem]{ulem}
\usepackage{subcaption} 
\usepackage{mathtools}
\theoremstyle{definition}

\begin{document}
\raggedbottom
\preprint{APS/123-QED}

\title{Stochastic Schr\"odinger Equations for Quantum Reverse Diffusion}

\author{Einar Gabbassov}
\email{egabbass@uwaterloo.ca}

\affiliation{Department of Applied Mathematics, University of Waterloo, Waterloo, ON N2L 3G1, Canada}

\affiliation{Perimeter Institute for Theoretical Physics, Waterloo, ON N2L 2Y5, Canada}

\affiliation{Institute for Quantum Computing, University of Waterloo, Waterloo, ON N2L 3G1, Canada}

\begin{abstract} The ensemble-averaged dynamics of open quantum systems are typically irreversible. We show that this irreversibility need not hold at the level of individually monitored quantum trajectories. Our main results are analytical stochastic Schr\"odinger equations for quantum reverse diffusion, along with corresponding stochastic master equations. These equations describe the exact and approximate stochastic reverse processes for continuously monitored Pauli channels, including time-dependent depolarizing noise. We show that the reverse processes generalize the forward dynamics by combining the noise effects of the forward processes with an additional stochastic drift that dynamically steers a quantum state back to its initial configuration. Consequently, the exact reverse stochastic Schr\"odinger equations admit closed-form solutions that can be implemented in real-time without the need for variational techniques. Our findings establish an analytical framework for quantum state recovery, noise-resilient quantum gates, quantum generative modelling, quantum tomography via forward-reverse cycles, and potential paradigms for quantum error correction based on reverse diffusion. \end{abstract}


\maketitle

\section{Introduction}
Reverse stochastic differential equations (SDEs) describe stochastic processes that undo statistical changes introduced in their respective forward stochastic processes \cite{anderson1982reverse,lindquist1979stochastic}. These equations form the theoretical backbone of modern classical generative diffusion models \cite{song2020score,yang2023diffusion,esser2024scaling,croitoru2023diffusion,liu2024sora,cao2024survey,kim2024diffusion}. Their core mechanism is to progressively degrade information in the data via a forward diffusion process and then learn reverse dynamics that restore it. This enables both the synthesis of new samples and the recovery of corrupted or missing information. The classical reverse SDEs, which govern the reverse diffusion, are highly nonlinear, as their drift depends on a probability density evaluated at the system's current state \cite{anderson1982reverse,song2020score}. This is incompatible with the quantum theory, where dynamics are fundamentally linear, and nonlinearities emerge only from post-measurement state normalization. In the quantum domain, numerous studies have proposed variational heuristics that train parameterized circuits to approximately simulate the reverse of a chosen noisy forward process \cite{zhang2024generative,kwun2025mixed,parigi2024quantum,hu2025local,liu2025measurement,cui2025quantum,zhang2025parameter}. One of the core assumptions of these techniques is that, during the reverse process, the effects of the original noise are absent, and a variational quantum circuit coherently imitates the reverse dynamics. Although powerful, these methods leave open the question of which physical principles fundamentally define a quantum reverse process, which, by definition, must incorporate the same noise and decoherence effects as the forward process \cite{anderson1982reverse,lindquist1979stochastic,song2020score}.

In this work, we bridge the theoretical gap by deriving the fully analytical quantum reverse stochastic Schr\"odinger equations for forward processes driven by measurement-induced Pauli noise, including time-dependent depolarizing noise. The derived reverse processes are generalizations of their forward counterparts; i.e., one can obtain the forward process from its reverse. Specifically, we show that the reverse processes are also a form of measurement-induced stochastic dynamics that incorporate a noise-aware stochastic drift. This drift, conditioned on the past and current measurement record, actively steers the quantum state back towards its initial configuration. Therefore, for the specified noisy forward processes acting on an arbitrary (possibly unknown) quantum state, we derive the reverse stochastic Schr\"odinger equations that define reverse dynamics that reconstruct the initial state under the same noise and decoherence as in the forward processes.

The presented reverse stochastic Schr\"odinger equations define an \textit{almost sure} reverse of the forward dynamics, in the sense that, conditioned on the measurement record, the state is driven back to its initial configuration with probability one. This yields significantly stronger convergence guarantees than the current variational heuristics, which by design converge in distribution. The stronger almost sure reversal of the forward dynamics can be relaxed by configuring the reverse process to steer the state onto a manifold of states; in this case, the dynamics implement a reversal in the distribution. Therefore, the reverse SDEs are quite powerful, as they enable a wide spectrum of applications from almost sure state recovery to quantum generative modelling.

In addition, we demonstrate that, unlike variational heuristics, the reverse processes can be implemented in real-time. Just as a forward process is naturally generated by the interaction with a monitored environment, the reverse process is naturally generated through the same kind of interaction, but with an additional feedback-controlled stochastic drift. This understanding enables the real-time implementation and alleviates the need for pre- or post-processing, local tomography, or offline variational techniques. Therefore, the reverse processes are not exclusive to being simulated with variational quantum circuits, but rather quantum phenomena that can arise in continuously monitored noisy systems with measurement-based feedback \cite{gross2018qubit,albarelli2024pedagogical,wiseman1994quantum,qi2009quantum}.
\begin{figure}[h]
    \centering
    \includegraphics[width=0.49\textwidth]{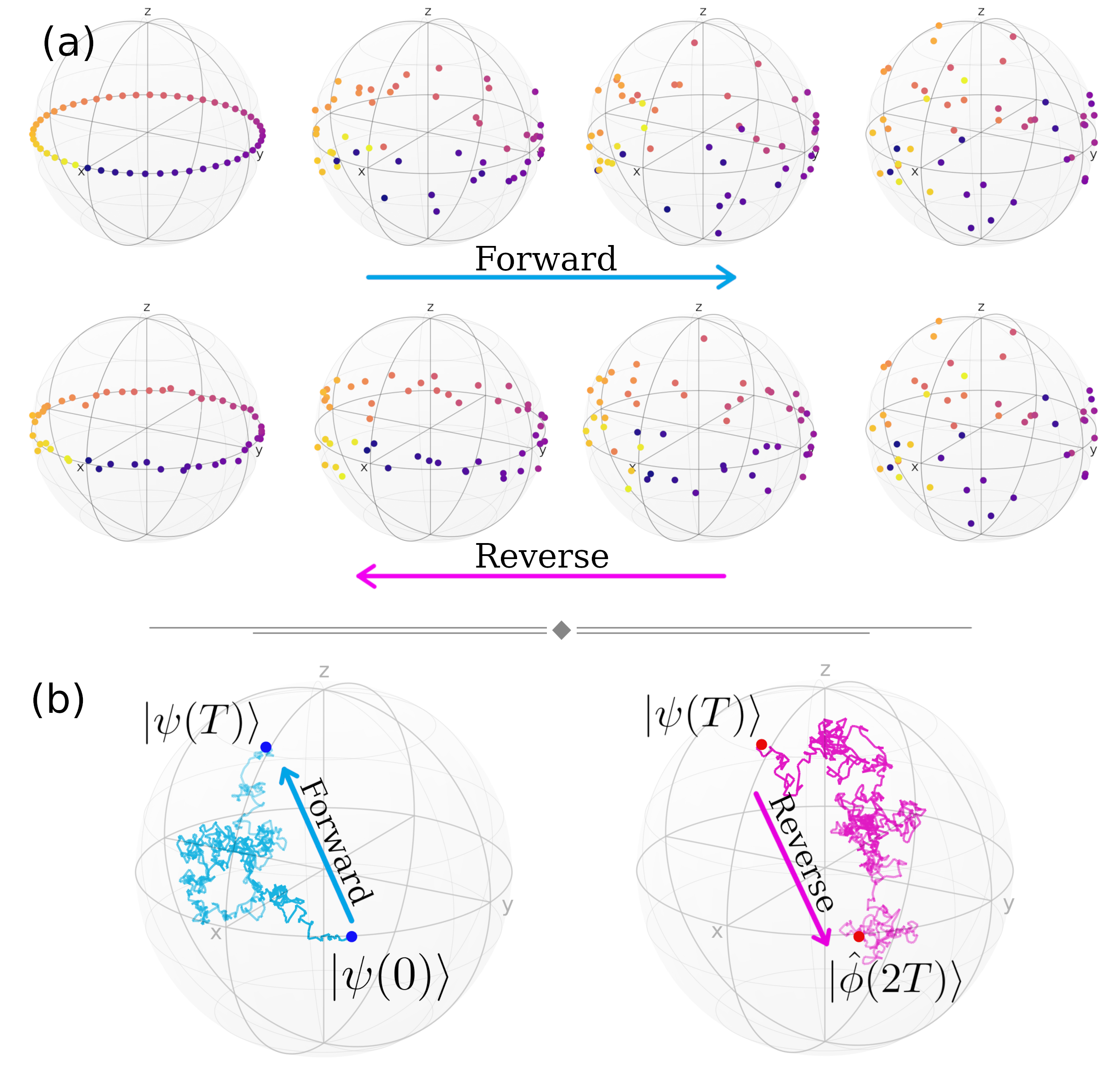}
    \caption{ (a) Forward and reverse depolarizing noise processes on an ensemble of states. Both processes occur under the same noise conditions, with the reverse process incorporating a noise-aware stochastic drift, which steers the states back to their initial configuration.
    (b) Individual quantum-state trajectories for the forward (blue) and corresponding reverse (purple) processes. The initial and final states of the forward process are $\ket{\psi(0)}$ and $\ket{\psi(T)}$, respectively. The reverse process starts at time $T$ and evolves $\ket{\psi(T)}$ into $|\hat \phi(2T)\rangle \approx \ket{\psi(0)}$.}
    \label{fig:fwd_rev_master}
\end{figure}

\section{Pauli Channels, Master Equations, and Stochastic Unravellings
\label{sec:pauli_channels_unravellings}}

In this section, we develop intuition for the relation between Pauli error channels, master equations, and individual quantum-state trajectories governed by stochastic Schr\"odinger equations. The central link is conditioning on the measurement record of a monitored environment. Discarding this record gives the ensemble-averaged formalism of channels and master equations. Conversely, retaining the record allows the evolution to be conditioned on the observed outcomes. This gives the pure-state trajectory formalism of stochastic Schr\"odinger equations.

Let $P \in \{\sigma_1,\sigma_2,\sigma_3\}^{\otimes m}$ be an $m$-qubit Pauli operator. A single Pauli error channel is the map
\begin{equation}\label{eq:single_pauli_channel}
    \rho \mapsto (1-q)\rho + q P\rho P ,
\end{equation}
where $q \in [0,1]$ is the probability that the Pauli error occurs. Thus, with probability $1-q$, the state $\rho$ is unchanged, and with probability $q$, the state is conjugated by $P$. For example, on a single qubit ($m=1$), different choices of $P$ correspond to bit-flip, bit-phase-flip or phase-flip errors. If the information about which error has occurred is not retained, then the output is the mixed state described by \cref{eq:single_pauli_channel}.

A time-dependent dynamical representation of the channel in \cref{eq:single_pauli_channel} is obtained by considering its action over a short time interval $\Delta t$ \cite{albarelli2024pedagogical,gross2018qubit}. If the error rate is $p$, then the probability of an error occurring during $\Delta t$ is approximately $p\Delta t$. Thus,
\begin{equation}\label{eq:density_state_update}
    \rho(t+\Delta t)
    \approx
    \left(1-p\Delta t\right)\rho(t)
    +
    p\Delta t\,P\rho(t)P.
\end{equation}
Subtracting $\rho(t)$ from both sides, dividing by $\Delta t$, and taking the limit $\Delta t \rightarrow 0$ yields the master equation
\begin{equation}\label{eq:single_pauli_master}
    \dot \rho = p(P\rho P-\rho).
\end{equation}
Physically, \cref{eq:single_pauli_master} describes the continuous-time limit of repeated system--environment interactions with discarded environment measurement outcomes.

Before introducing stochastic Schr\"odinger equations, which provide a pure-state trajectory formalism, we illustrate their relation to the ensemble-averaged formalism. The key distinction is whether the measurement record of the environment is discarded or retained. Let $\ket{\psi(t)}_S$ and $\ket{\phi}_E = \alpha\ket{0}_E+\beta\ket{1}_E$ be single qubit states of a system $S$ and environment $E$, respectively. Suppose that, during the interval $[t,t+dt)$, $E$ interacts with $S$ through a controlled-$P$ gate, with $E$ as the control and $S$ as the target. The joint state after the interaction is
\begin{align}\label{eq:joint_sys_env_state_example}
    \ket{\Psi(t+dt)}_{ES} =\alpha\ket{0}_E\ket{\psi(t)}_S + \beta\ket{1}_E P\ket{\psi(t)}_S .
\end{align}
If the environment is measured in the computational basis and the outcome is discarded, the system is described by the unconditional state
\begin{align*}
    \rho_S(t+dt) &= \Tr_E \ket{\Psi(t+dt)}\bra{\Psi(t+dt)}_{ES} \\
    &= |\alpha|^2\rho_S(t) + |\beta|^2 P\rho_S(t)P \\
    & = (1 - |\beta|^2) \rho_S(t) + |\beta|^2 P\rho_S(t)P,
\end{align*}
where $\rho_S(t)=\ket{\psi(t)}\bra{\psi(t)}_S$. Thus, the unconditional description is an ensemble average over the possible measurement outcomes. Note the structural similarity to \cref{eq:density_state_update}

If the measurement outcome is retained, the system state is conditioned on the observed value of the environment qubit. Inspecting the two terms in \cref{eq:joint_sys_env_state_example},
\begin{align*}
    \text{outcome } \ket{0}_E
    &\implies
    \ket{\psi(t+dt)}_S=\ket{\psi(t)}_S,\\
    \text{outcome } \ket{1}_E
    &\implies
    \ket{\psi(t+dt)}_S=P\ket{\psi(t)}_S,
\end{align*}
with probabilities $|\alpha|^2$ and $|\beta|^2$. Encoding this measurement record with the binary stochastic increment $dN$, with $dN=0$ for outcome $0$ and $dN=1$ for outcome $1$, gives
\begin{equation*}
    \ket{\psi(t+dt)}_S
    =
    \bigl(1-dN\bigr)\ket{\psi(t)}_S
    +
    dN \, P\ket{\psi(t)}_S .
\end{equation*}
Subtracting $\ket{\psi(t)}_S$ from both sides gives the stochastic differential form describing the dynamics of a pure state:
\begin{equation}\label{eq:simple_jump_equation}
d\ket{\psi(t)}_S = \big (-I \, dN + P \, dN \big )\ket{\psi(t)}_S
\end{equation}
Here, $d\ket{\psi(t)}_S=\ket{\psi(t+dt)}_S-\ket{\psi(t)}_S$. Since $dN\in\{0,1\}$, \cref{eq:simple_jump_equation} selects one of the two normalized conditional updates above: the state is unchanged for $dN=0$, or the unitary $P$ is applied for $dN=1$. Thus, the same interaction gives either an ensemble-averaged mixed-state update when the outcome is discarded or a pure stochastic trajectory when the outcome is retained.

Such a trajectory-level description is called \textit{stochastic unravelling}. This unravelling is essential when the state of a single realization matters. For example, in monitored quantum systems or quantum computations, one acts on the state produced in the current run rather than on an ensemble average. At the same time, the master equation is readily recovered by averaging over the conditioned trajectories.

\subsection*{Forward Process}
The controlled-$P$ example model leading to \cref{eq:simple_jump_equation} gives a counting-type stochastic update, where the retained measurement outcome is encoded in the binary increment $dN$. In a continuous weak-measurement scheme, the system instead interacts weakly with the environment over each interval $dt$, after which the environment is measured. This produces a continuous diffusive measurement record, modelled with a Wiener increment $dW$.

To describe the evolution under this continuous record, we define the \textit{forward process} as the diffusive unravelling of the master equation in \cref{eq:single_pauli_master}. It is described by the stochastic Schr\"odinger equation \cite[Sec. 2.3]{barchielli2009quantum}:
\begin{align}\label{eq:simple_fwd_sde}
    &d\ket{\psi} = \left( -\frac{p}{2} I dt + \sqrt{p} \, L \, dW \right)\ket{\psi(t)},  \nonumber \\
    &\ket{\psi(0)} = \ket{\psi_0}, \quad \quad 0 \leq t \leq T 
\end{align}
Here, the jump operator $L$ is defined as $L = P$ for the \textit{information-dissipative} case, or $L = iP$ for the \textit{information-conserving} case. The constant $p \in [0,1]$ is the noise strength, and $dW$ is an observed stochastic increment satisfying $(dW)^2 = dt$. The stochastic term $\sqrt{p} L dW$ represents the measurement-induced stochastic perturbation, while the deterministic term $-\frac{p}{2} I dt$ accounts for the resulting measurement backaction.

The information-dissipative case describes the leakage of information into the environment and, subsequently, into the measurement record. This makes the evolution dissipative and hence non-unitary. The unit norm state $|\hat \psi(t)\rangle$ (a posteriori state) can be obtained by normalization \cite[Sec. 2.4]{barchielli2009quantum},
\begin{equation}\label{eq:normalized_fwd_state}
    |\hat \psi(t) \rangle = \frac{\ket{\psi(t)}}{\|\ket{\psi(t)}\|}.
\end{equation}
The observed stochastic increment is
\begin{equation}\label{eq:observed_increment}
    dW = \sqrt{p} \, \langle L + L^{\dagger}\rangle_{\hat \psi_t} dt + d\hat W. 
\end{equation}
This increment carries a state-dependent signal through the expectation value 
$$\langle L+L^{\dagger}\rangle_{\hat \psi_t}:= \langle \hat \psi(t)|L+L^{\dagger}|\hat \psi(t)\rangle,$$
which is subject to noise given by the standard Wiener increment $d\hat W$.
Increasing $p$ captures the fundamental trade-off of measurement: the more information is extracted from the record in \cref{eq:observed_increment}, the more the state is perturbed in \cref{eq:simple_fwd_sde}.

\begin{figure}[t]
    \centering
    \includegraphics[width=0.33\textwidth]{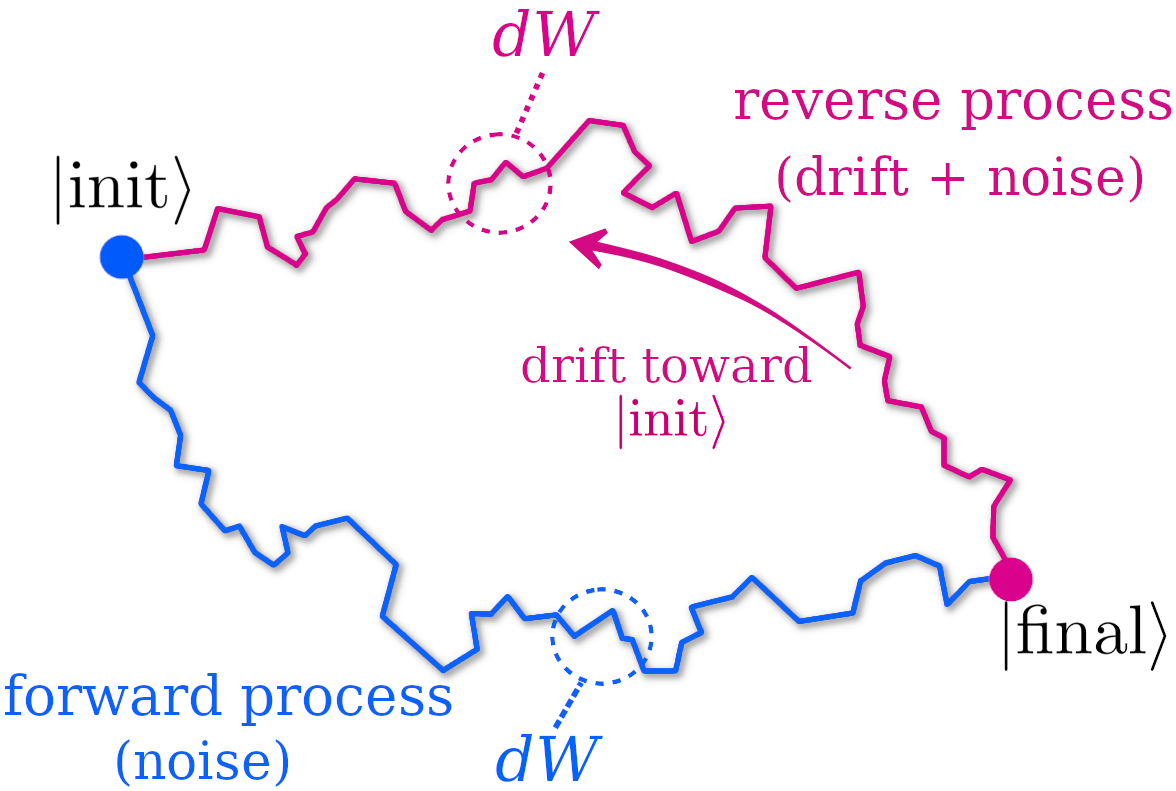}
    \caption{Schematic representation of the forward and reverse trajectories in pure-state space. The forward process evolves the state from $\ket{\mathrm{init}}$ to $\ket{\mathrm{final}}$, tracing a stochastic path under perturbations from the noise increments $dW$ (blue). Analogous to a particle undergoing Brownian motion in a medium, the forward trajectory fluctuates randomly without a preferred direction in state space. The subsequent reverse process starts from $\ket{\mathrm{final}}$ and traces a generally different stochastic path (pink). It remains stochastic, but includes an additional drift that steers the state back to its initial configuration $\ket{\mathrm{init}}$.}
    \label{fig:concept_fwd_rev}
\end{figure}

Conversely, in the information-conserving case, the evolution is unitary, and $L+L^{\dagger}=0$. Hence, $dW = d\hat W$ is a standard Wiener increment that carries no information about $|\hat \psi(t)\rangle$, thereby conserving information within the system. This case corresponds to random unitary rotations generated by the Pauli operator $P$.

\subsection*{Conceptual Overview of Reverse Process}
The trajectory-level description in \cref{eq:simple_fwd_sde} is the starting point for the stochastic Schr\"odinger equation that describes the reverse dynamics presented in the next section. The reverse process evolves under the same type of noise as the forward process, but with an additional stochastic drift that steers the state toward its initial configuration. Conceptually, the forward process generates a stochastic path in pure-state space from the initial state to a final noisy state.  The reverse process starts from this final state and follows a generally different stochastic path back. Because the reversal is formulated at the level of individual trajectories, the normalized terminal state is recovered exactly, rather than only in the ensemble average; see the conceptual schematic in \Cref{fig:concept_fwd_rev}.

\section{Results \label{sec:results}}
Pauli error channels are known to be non-invertible. Their continuous-time version, given by the Lindblad equations, is also known to be irreversible. We show that this is not the case if we consider their stochastic unravellings. Working at the level of the actual continuous-time trajectory of an individual quantum state, rather than the ensemble average, we construct a stochastic process that reverses the effects of Pauli noise. Therefore, for a quantum state that is continuously perturbed by random rotations or weak measurements, we demonstrate a reverse process that, under unit measurement efficiency and instantaneous feedback, recovers the initial state. The recovery is exact for single Pauli error channels and approximate for multiple Pauli error channels, e.g., depolarizing noise (\Cref{sec:applications}). We remark that the recovery is happening under the same noise effects as in the forward process. In \mbox{\Cref{fig:fwd_rev_master}}, we demonstrate the dynamics of the forward and reverse processes for depolarizing noise. Finally, in \Cref{sec:imperfect_measurement_delayed_feedback}, we extend the reverse process to include finite measurement efficiency and feedback delay, and numerically characterize their effect on state recovery.

\subsection*{Reverse Process}
The initial state $\ket{\psi_0}$ that has undergone the forward process in \cref{eq:simple_fwd_sde} for a duration of time $T$ can be recovered exactly using a subsequent reverse process of the same duration $T$. For $T \leq t \leq 2T$, the following stochastic Schr\"odinger equation describes the dynamics of the reverse process:
\begin{align}\label{eq:simple_rev_sde}
    d\ket{\phi(t)}
    &=
    \left( \left( -\frac{p}{2}I - \frac{X(t)}{2T-t}\sqrt{p}L \right)dt + \sqrt{p}\,L\,dW \right) \ket{\phi(t)}, \nonumber \\
    dX(t)
    &=- \frac{X(t)}{2T-t}dt + dW, \quad T \leq t \leq 2T
\end{align}
In the above, $X(t)$ denotes a scalar process, and
\begin{equation*}
    dW
    =
    \sqrt{p}\,
    \langle L+L^{\dagger} \rangle_{\hat \phi_t}dt
    +
    d\hat W
\end{equation*}
carries information about the reverse state
\begin{equation*}
    |\hat \phi(t) \rangle
    =
    \frac{\ket{\phi(t)}}{\|\ket{\phi(t)}\|}.
\end{equation*}
The initial conditions for the reverse process are given by the terminal state of the forward process and the terminal value of the measurement record:
\begin{equation*}
    |\phi(T)\rangle
    =
    |\hat \psi(T) \rangle,
    \qquad
    X(T)
    =
    W(T)
\end{equation*}
The normalized terminal state $|\hat \phi(2T)\rangle$ of the reverse process converges exactly to the initial state $\ket{\psi_0}$ of the forward process, that is,
\begin{equation*}
    \left|\langle \psi_0 \, | \, \hat{\phi}(2T)\rangle\right|^2 = 1 .
\end{equation*}
Compared with the forward SDE in \cref{eq:simple_fwd_sde}, the reverse SDE contains the same noise term $\sqrt{p}L\,dW$, but also an additional drift term proportional to $X(t)/(2T-t)$. Furthermore, the reverse process is the statistical time-reversal of the forward SDE: the distribution of $\ket{\phi(t)}$ for $t \in [T, 2T]$ coincides with the time-reversed distribution of $\ket{\psi(t)}$ for $t \in [0, T]$.

\subsubsection*{Reverse Stochastic Master Equation}
Using \cref{eq:simple_rev_sde} and It\^o calculus, it is straightforward to derive the reverse stochastic master equation (SME). For $T \leq t \leq 2T$, we have
\begin{align}\label{eq:sme_simple_rev}
    d\tilde{\rho}&= \big( \mathcal{K}(t,\tilde \rho) + \mathcal{L}(\tilde \rho) \big ) \, dt + \sqrt{p} \{L,\tilde{\rho}\} \, dW,
\end{align}
where
\begin{align*}
    \mathcal{L}(\rho) &= p(L\rho L^{\dagger} - \rho), \\
    \mathcal{K}(t, \rho) &= -\frac{X(t)}{2T-t}\sqrt{p} \, \{L, \rho\},\\
    \{L,\rho\} &= L\rho+\rho L^{\dagger}.
\end{align*}
The superoperator $\mathcal{L}$ is the Lindbladian of the Pauli error channel. For the information-conserving case, $L=iP$ and the reverse SME becomes
\begin{align}\label{eq:sme_simple_rev_inf_conserv}
    d\tilde{\rho}
    =
    \left(
        -i\frac{X(t)}{2T-t}\sqrt{p}
        \left[P,\tilde{\rho}\right]
        +
        \mathcal{L}(\tilde{\rho})
    \right)dt
    +
    i\sqrt{p}
    \left[P,\tilde{\rho}\right]dW.
\end{align}
We note that removing the terms proportional to $X(t)$ removes the contribution of the reverse drift and reduces \cref{eq:simple_rev_sde,eq:sme_simple_rev,eq:sme_simple_rev_inf_conserv} to the diffusive unravelling in \cref{eq:simple_fwd_sde} of the forward master equation in \cref{eq:single_pauli_master}. Furthermore, the singularity at $t=2T$ is integrable due to the Brownian bridge property, $X(2T)=0$. Thus, the total action $\int_T^{2T} X(t)\sqrt{p}\{L,\tilde \rho\}/(2T-t)dt$ is finite almost surely.

\subsection*{Intuition}
Let us develop intuition for the simple reverse SDEs in \cref{eq:simple_rev_sde}. These insights directly carry over to reverse SMEs. The reverse SDEs do not contain or require any information about the initial state we want to recover. Indeed, the reverse process operates ``blindly" on any quantum state that has undergone the forward process for time $T$ with a measurement record $W(T)$. Mathematically, this is reflected in the fact that the reverse SDE in \cref{eq:simple_rev_sde} does not depend on the initial state $\ket{\psi_0}$ or any intermediate state $\ket{\psi(t)}$ of the forward process.

Furthermore, the reverse SDE operates under the same noisy conditions as the forward process, as evidenced by the noise terms $\sqrt{p} LdW$. Unlike the forward SDE in \cref{eq:simple_fwd_sde}, its reverse counterpart features a stochastic drift $-X(t)\sqrt{p}L/(2T-t)dt$, where $X(t)$ is a Brownian bridge which contains the memory of the forward process. The drift dynamically drives the quantum state toward its initial configuration $\ket{\psi_0}$. Both the drift and the noise term ensure that the reverse process is a statistical reverse of the forward process. However, if we remove the drift term, then we recover the forward process.

For the information-conserving reverse SDE, the drift term can be identified with a Hamiltonian $H(t)=X(t)\sqrt{p} P /(2T-t)$ multiplied by $-i$, such that $-iH(t)dt = -iX(t)\sqrt{p} P /(2T-t)dt$. It follows that $H(t)$ generates a unitary evolution. In the information-dissipative case, the drift term corresponds to the imaginary time evolution because $-iH(t)dt = -X(t)/(2T-t)\sqrt{p}P dt$. The terms proportional to the identity in both SDEs are the It\^o correction terms, often interpreted as measurement backaction.

\begin{figure}[t]
    \centering
    \includegraphics[width=0.49\textwidth]{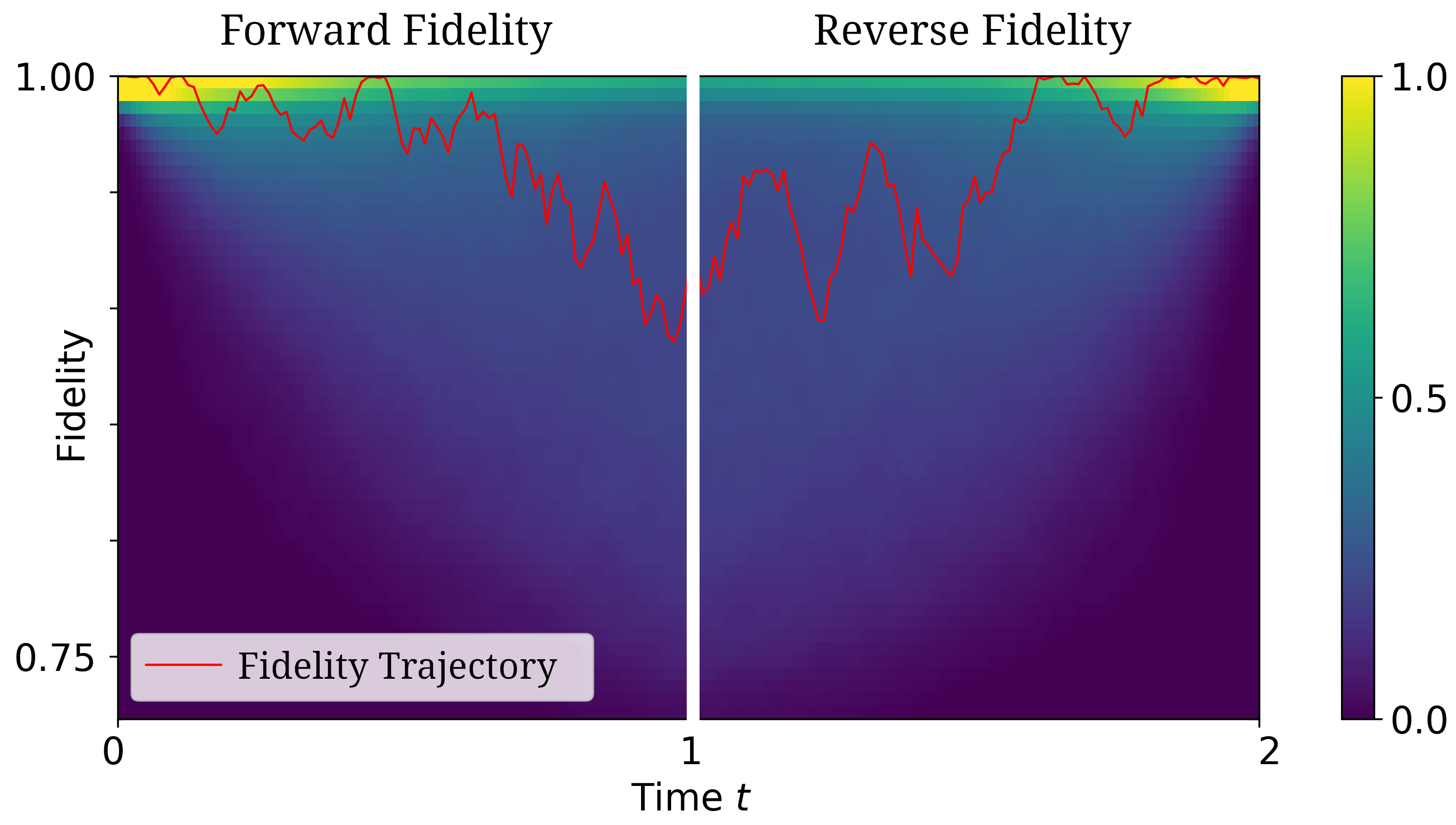}
    \caption{Quantum-state fidelity probability flow under forward and reverse dynamics, with a representative single-trajectory realization shown in red. Here, $L=\sigma_x$ and $p=0.2$. The forward and reverse processes occur on the time intervals $[0,1]$ and $[1,2]$, respectively. During the forward process, the fidelity distribution spreads away from unity. During the reverse process, it reconcentrates at unity, indicating recovery of the initial state.}
    \label{fig:fidelity_plot}
\end{figure}

In \Cref{fig:fidelity_plot}, we demonstrate the probability flow of the fidelity between an initial state $\ket{\psi_0}$ and its forward and then reversed stochastic states. In the forward segment, fidelities typically diffuse away from unity, indicating a progressive loss of the overlap with the initial state. In the reverse segment, the fidelity distribution reconcentrates around unity, reflecting recovery of the initial state.

\section{Methods}\label{sec:methods}
In this section, we discuss how to construct the reverse SDE in \cref{eq:simple_rev_sde}. We assume the forward process runs for a total duration of $2T$, with the reverse process activating halfway through at $t=T$, while the forward process remains active. Given the initial state $\ket{\psi_0}$, the solution of the forward process in \cref{eq:simple_fwd_sde} is $\ket{\psi(t)} = F(t)\ket{\psi_0}$ where $F(t)$ is defined as:
\begin{equation}
    F(t) := \exp\left ( -p t + \sqrt{p} L W(t) \right), \ 0 \leq t \leq 2T
\end{equation}
For the reverse process, we define another operator:
\begin{align}\label{eq:reverse_operator_simple}
    R(t):=
    \begin{cases}
        I,  \quad 0 \leq t < T, \\
        \exp \Big(\left( X(t) - W(t) \right)\sqrt{p} L \Big ), \ T \leq t \leq 2T
    \end{cases}
\end{align}
The operator above is the identity up until time $T$, and $X(T) = W(T)$. Then, for $T \leq t \leq 2T$, the solution to the reverse process in \cref{eq:simple_rev_sde} is given by:
\begin{equation}\label{eq:solution_to_reverse_simple}
    \ket{\phi(t)}=R(t)F(t)\ket{\psi_0},\ T \leq t \leq 2T
\end{equation}
We note that $F(t)$ induces noise for the entire duration, while $R(t)$, starting from $t=T$, uses this noise to drive the state to its initial configuration by the time $t=2T$. Specifically,
\begin{equation*}
    |\hat \phi(2T) \rangle = \frac{R(2T)F(2T)\ket{\psi_0}}{\| R(2T)F(2T)\ket{\psi_0} \|} = \ket{\psi_0}. 
\end{equation*}
In the information-dissipative case, neither operator preserves the norm, but this is not an issue as post-normalization can always be applied \cite{barchielli2009quantum} without breaking the analysis. The information-conserving reverse processes are realized via a coherent drift Hamiltonian that integrates the increments $dX(t)$. This means that such processes can be implemented in situ via real-time unitary feedback. In contrast, implementing information-dissipative dynamics is significantly more challenging, as it corresponds to imaginary time evolution (ITE). Conventional ITE algorithms typically require local tomography of the quantum state at each time step and extensive offline processing \cite{motta2020determining,mcardle2019variational,lin2021real,kolotouros2025accelerating,mittal2025deterministic,ray2025quasiprobabilistic,rrapaj2025exact}. In our setup, the instantaneous ITE drift provides real-time feedback that incorporates the continuous stream of stochastic measurement increments $dW$ and assumes no knowledge of the quantum state. Therefore, the existing ITE methods are not suitable for our setup. An alternative approach could be unitary block encoding routines \cite{vasconcelos2025methods,low2019hamiltonian,berry2015hamiltonian,berry2015simulating,shang2025designing}. For example, it is possible to block encode scaled-down $R(t)/\alpha(t)$ into a larger unitary operator acting on an ancilla-dilated system. Measuring the ancilla and post-selecting yields $R(t)/\alpha(t)\ket{\phi(t)}$. However, realizing the desired outcome near-deterministically (with controlled probability of success) is impossible. The techniques, such as oblivious amplitude amplification \cite{yan2022fixed,guerreschi2019repeat,berry2015hamiltonian,berry2015simulating}, will not work because, as they amplify the probability of measuring the right outcome, they inevitably implement a polynomial transformation of $R(t)/\alpha(t)$, which significantly distorts the desired dynamics \cite{zecchi2025improved,guerreschi2019repeat}. Consequently, realizing online, near-deterministic, information-dissipative reverse dynamics requires a different strategy. In \Cref{sec:algo}, we address all these challenges and demonstrate that the reverse process can be realized through a series of weak measurements and state teleportations, during which the state $|\hat \phi(t)\rangle$ effectively realizes the reverse dynamics. Additionally, in \Cref{sec:ap_resource_analysis}, we perform a resource analysis of the algorithm and show that the fully real-time, near-deterministic implementation does not incur infeasible resource overhead. Below, we present an outline of the algorithm.

\subsection*{Algorithm Outline for Dissipative Case}
The representation in \cref{eq:solution_to_reverse_simple} expresses the reverse trajectory as the combined action of the continued noisy evolution $F(t)$ and the reverse operator $R(t)$. The operator $F(t)$ accounts for the same stochastic evolution that appears in the forward process, while $R(t)$ supplies the additional drift that steers the state back toward its initial configuration. This decomposition is directly useful for implementation because it isolates the part of the dynamics that must be added to the continuously monitored noisy evolution.

The implementation challenge is therefore to realize a normalized reverse operator in real time without knowledge of the current system state. Applying this operator directly to the system would require post-selection. An unsuccessful post-selection event would then irreversibly corrupt the state and halt the reverse process. To avoid this, the normalized reverse operator is encoded into independent auxiliary resource states.

Specifically, at each time step $\Delta t$, duplicate resource states are prepared, each carrying the normalized version of the incremental reverse operator. One such resource state is coupled to the system through a gate-teleportation procedure \cite{bennett1993teleporting,gottesman1999demonstrating}, which transfers the incremental reverse operator onto the quantum state. Since a single transfer attempt is probabilistic, the procedure is repeated with additional resource states, resulting in a near-deterministic implementation with a controllable failure probability. Full implementation details and resource estimates are given in \Cref{sec:algo}.

\section{Applications\label{sec:applications}}
Reverse SDEs are interesting mathematical objects that provide alternative perspectives on noisy processes and can serve as a foundation for constructing new stochastic processes and quantum applications. Below, we present several examples.

\subsection*{Reverse Depolarizing Noise}
Equipped with insights from the exact reverse SDEs for single Pauli error channels in \cref{eq:simple_rev_sde}, we can construct the approximate reverse stochastic Schr\"odinger equation for multi-Pauli-error channels, such as depolarizing noise. See \Cref{fig:fwd_rev_master} for the demonstration. Let us consider the diffusive unravelling \cite{jacobs2006straightforward} of the depolarizing noise, which we refer to as a forward process. The dynamics of this process are described by the stochastic Schr\"odinger equation:
\begin{align}\label{eq:forward_dn_dissipative}
    d\ket{\psi(t)} &= \left(-\frac{1}{2}\sum_{k=1}^3 \frac{p}{3}L_k^{\dagger}L_k dt + \sum_{k=1}^3\sqrt{\frac{p}{3}}L_k dW_k \right) \ket{\psi(t)} \nonumber \\
    \ket{\psi(0)} &= \ket{\psi_0}, \ \  0 \leq t \leq T
\end{align}
We note that there are three distinct non-commuting error channels, each manifesting as a stochastic term $\sqrt{p/3} \, L_k \, dW_k(t)$, 
where $L_k$ are defined as $L_k = \sigma_k$ (information-dissipative case), or $L_k = i\sigma_k$ (information-conserving case). The stochastic increments $dW_k$ satisfy $(dW_k)^2 = dt$ and $dW_k dW_j =0$ for $k \neq j$.

The exact reverse SDE for depolarizing noise admits no closed-form construction; it entails an infinite Magnus (nested-commutator) series coupled to a countably infinite nonlinear auxiliary hierarchy of scalar SDEs. However, despite these challenges, we give an approximate reverse SDE.

The SDEs in \cref{eq:reverse_sdes} define a reverse process $\ket{\phi(t)}$ that starts from $\ket{\phi(T)}=| \hat \psi(T) \rangle$ and continuously evolves toward $\ket{\psi_0}$ over the interval $[T,2T]$ while all error channels (noise) remain active. In the regime $pT < 1$, there exist a constant $c >0$ such that the normalized terminal state $|\hat{\phi}(2T)\rangle:=\ket{\phi(2T)}/\|\ket{\phi(2T)}\|$ approximates the target state $\ket{\psi_0}$ with the expected fidelity 
\begin{equation}
    \mathbb{E} \left [\  \text{Fidel} \big ( |\hat \phi(2T)\rangle, \ket{\psi_0} \big ) \right ] \geq 1 -  c \left ( p T \right)^3.
\end{equation}
For $T \leq t \leq 2T$, the following SDEs describe the dynamics of such a process:
\begin{align}\label{eq:reverse_sdes}
    d\ket{\phi(t)}
    &=\left(\mathcal{D}(t) dt + \sum_{k=1}^3 \mathcal{H}_k(t) dX_k\right) \ket{\phi(t)}, \nonumber \\
    dX_k &= -\frac{X_k(t)}{2T-t}dt + \gamma dW_k, \quad  T \leq t \leq 2T
\end{align}
In the above, $X_k(t)$, for $k=1,2,3$, are scalar processes. The operators $\mathcal{D}(t)$ and $\mathcal{H}_k(t)$, and the complex constant $\gamma$, are defined as:
\begin{align}\label{eq:reverse_operators_disspat}
    \mathcal{D}(t) &:= -p I + \frac{\gamma^2}{2}\left(3 - 2\sum_{k=1}^3 (X_k(t)-X_k(T))^2 \right ) I, \nonumber \\
    \mathcal{H}_k(t) &:= \sigma_k + \frac{1}{2}\sum_{j=1}^3[\sigma_j,\sigma_k] \left(X_j(t)-X_j(T) \right),\nonumber \\
    \gamma &:= \sqrt{\frac{p}{3}} + 2i\frac{p}{3}
\end{align}
Similarly, for the information-conserving forward process, its reverse follows the same structure as given in \cref{eq:reverse_sdes}, but with the following operators and constant:
\begin{align}\label{eq:reverse_operators_conserv}
    \mathcal{D}'(t) &:= -\frac{\gamma'^2}{2}\left(3 + 2\sum_{k=1}^3 (X_k(t)-X_k(T))^2 \right ) I, \nonumber \\
    \mathcal{H}'_k(t) &:= i \left( \sigma_k + \frac{i}{2}\sum_{j=1}^3[\sigma_j,\sigma_k] (X_j(t)-X_j(T)) \right), \nonumber \\
    \gamma' &:= \sqrt{\frac{p}{3}} -2\frac{p}{3}
\end{align}
The initial conditions for the scalar processes $X_k(t)$ are determined by integrating the observed increments from the forward process; i.e.,  $$X_k(T) = \sqrt{\frac{p}{3}}W_k(T) + 2( i )^{b} \, \frac{p}{3} \, S_k(T),$$
where $b=1$ for the information-dissipative case, and $b=0$ for the information-conserving case. Here, 
\begin{align*}
    S_1(T) &:= 1/2\int_0^T (W_2dW_3-W_3dW_2),\\
    S_2(T) &:= 1/2\int_0^T (W_3dW_1-W_1dW_3),\\
    S_3(T) &:= 1/2\int_0^T (W_1dW_2-W_2dW_1)
\end{align*}
are known as L\'evy's stochastic areas \cite{helmes1983levy}. These stochastic quantities are easily reconstructed from the observed records $W_k$. In a physical sense, $S_k(T)$ captures pairwise interactions between noise perturbations given by $L_n dW_n$ and $L_m dW_m$ in \cref{eq:forward_dn_dissipative}. See the construction of the reverse SDE in \Cref{sec:app_rev_dn}.

\subsection*{Diffusion-Driven Quantum Gates}
The reverse processes introduced in \cref{eq:simple_rev_sde} can be extended to a framework for generating quantum gates driven by diffusion. Consequently, this enables the diffusion-based generation of quantum states. Let $P = A \otimes B$ with $A,B \in \{I, \sigma_1, \sigma_2, \sigma_3\}$. Conventionally, a single- or two-qubit gate $G(\theta) := \cos(\theta)I -i\sin(\theta)P$ are achieved via coherent evolution governed by the Schr\"odinger equation $d\ket{\psi}/dt = -iH\ket{\psi(t)}$, where $H := P$. However, when subjected to single-channel noise, the dynamics are instead governed by a forward SDE:
$$d\ket{\psi} = \left( -iH dt -\frac{p}{2} I dt + i\sqrt{p} P dW \right)\ket{\psi(t)},$$
where $dW$ is a standard Wiener increment. Despite this being an information-conserving process, the noise prevents the deterministic implementation of $G(\theta)\ket{\psi_0}$. We show that instead of trying to cancel or correct the noise, we can use it to drive the system from its initial state $\ket{\psi_0}$ to a desired target state $G(\theta)\ket{\psi_0}$. Assuming the noise strength $p$ is known, and $\theta$ is a target rotation angle, the SDEs generating such gates are:
\begin{align}\label{eq:noise_absorbing_sdes}
    d\ket{\psi(t)} &= \left(\mathcal{D}' dt + i\sqrt{p}P dW \right)\ket{\psi(t)}, \nonumber \\
    \mathcal{D}' &:= -\frac{p}{2}I -i \frac{\theta/\sqrt{p} + X(t)}{2T-t}\sqrt{p}P, \nonumber \\
    dX &= -\frac{\theta/\sqrt{p} + X(t)}{2T-t}dt + dW, \ X(T)=0
\end{align}
At $t=2T$, the SDE deterministically yields $\ket{\psi(2T)} = G(\theta)\ket{\psi_0}$. The singularity in the drift at $t=2T$ is integrable because $X(2T) = -\theta/\sqrt{p}$. Furthermore, while the final time $2T$ can be freely chosen, a smaller $T$ implies faster evolution, requiring a higher instantaneous Hamiltonian strength. The implementation of such a diffusion process is straightforward. We identify the drift Hamiltonian to be
\begin{equation}\label{eq:diffusion_gates_driver}
    \hat H(t) :=\frac{\theta/\sqrt{p}+X(t)}{2T-t}\sqrt{p} \, P,
\end{equation}
where $X(t)$ is constructed from the observed increments $dW$ according to \cref{eq:noise_absorbing_sdes}. Then, the SDE
$$d\ket{\psi} = \left( -i \hat H(t) dt -\frac{p}{2} I dt + i\sqrt{p} P dW \right)\ket{\psi(t)},$$ implements $G(\theta)\ket{\psi_0}$. 
The equation above shows that noise-resilient gates can be implemented by a simple coherent feedback Hamiltonian in \cref{eq:diffusion_gates_driver}. Furthermore, we can promote $\theta$ to a random variable drawn from a specified distribution, so that the dynamics drive the state back not to a single target, but to a manifold of states.

\subsection*{Quantum Tomography}
Quantum state tomography aims to reconstruct an unknown quantum state. An accurate reconstruction generally requires numerous identical copies of the state. Each copy is measured and hence perturbed. Depending on the measurement strength, the copy can be slightly perturbed or destroyed entirely \cite{monroe2021weak}. The reverse SDE and reverse SME in \cref{eq:simple_rev_sde,eq:sme_simple_rev} suggest that, at least in principle, it is possible to supplement the weak-measurement tomographic protocol \cite{wu2013state,hofmann2010complete} on $\ket{\psi_0}$ with an additional round of measurements performed while the state is being driven back toward its initial configuration. To illustrate this, we subject a single copy to a continuous weak measurement of $P$ with measurement strength parameter $p$ and duration $T$. This process yields a forward information-dissipative evolution described by \cref{eq:simple_fwd_sde} with $L = P$. In this case, the observed measurement increment is $dW = 2\sqrt{p}\,\langle P \rangle_{\hat \psi_t}\,dt + d\hat W$, where $2\sqrt{p}\,\langle P \rangle_{\hat \psi_t}\,dt$ is the useful signal with the error $d\hat W$. We can immediately see that the parameter $p$ balances a trade-off between the signal strength and the amount of perturbation introduced into the system. After the forward process, we implement the reverse diffusion process (as in \cref{eq:simple_rev_sde} or \cref{eq:sme_simple_rev}) on the interval $[T,2T]$ driven by the same type of continuous measurements. The process produces a stochastic path $|\hat \phi(t)\rangle$ that has the same statistics as the forward path, but in reverse. After the duration $T$, the normalized reverse process converges to the unknown initial state $\ket{\psi_0}$. This opens up the possibility of performing tomography not only during the forward weak-measurement phase, but also during the reverse phase, effectively providing two measurement stages on the same physical state within a single forward–reverse cycle. By iterating such forward–reverse experiments over many copies, we accumulate an ensemble of trajectories $W(t)$. Subsequently, one can process the aggregated data with quantum filtering (Belavkin–SME) \cite{bouten2007introduction,liang2025stabilization,gough2012principles} and perform Bayesian or maximum-likelihood quantum state estimation \cite{lukens2020practical,ramoa2025bayesian,hsieh2025bayesian,clark2025efficient}.

\subsection*{Correcting Errors} We examine a scenario where a quantum system experiences single-qubit Pauli errors induced by its continuously monitored environment. Conceptually, this involves a quantum system interacting with a continuous stream of ancillas, representing a memoryless environment. Each ancilla briefly interacts with the system before being measured, and the system then engages the next ancilla. Due to entanglement, the ancilla's measurement leaves a corresponding imprint (perturbation) on the system's state. In the continuous limit, this process yields a continuous measurement record with increments $dW = \sqrt{p} \,\langle L+L^{\dagger} \rangle_{\hat \psi_t} dt + d\hat W$. Since the environment constitutes part of the measurement apparatus, the errors it induces onto the state are known as measurement-induced errors \cite{ahn2003quantum}. Consequently, the quantum reverse SDEs suggest that instead of correcting errors as they appear, we can let the errors accumulate for some time $T$ and then implement the reverse process given in \cref{eq:simple_rev_sde,eq:reverse_sdes}, thereby cancelling all current and accumulated Pauli errors. This is especially beneficial when $L + L^{\dagger} = 0$, as the reverse dynamics can be implemented through a coherent drive.

\section{Imperfect Measurement Efficiency and Feedback Delay}\label{sec:imperfect_measurement_delayed_feedback}
In previous sections, we have established the stochastic Schr\"odinger equations for reverse diffusion. In practice, the measurement record used to construct the reverse process is subject to finite efficiency and feedback latency. Here, we extend the equations to incorporate the effects of imperfect measurement and feedback delay on the trajectory level. Furthermore, we qualitatively characterize their impact on state recovery.

Imperfect measurement efficiency means that the measurement record does not contain the full information about the system-environment interaction. Part of this information is lost and is therefore not recorded. This unobserved information leads to a loss of purity, so the conditioned state generally becomes mixed. At the trajectory level, we represent this incomplete information by an experimentally accessible measurement record that is only partially correlated with the true increment driving the underlying stochastic evolution. We model this by replacing the ideal (complete) measurement increment $dW$ defined in \cref{eq:observed_increment} with the observed inefficient increment $dU$. This increment contains only partial information, and it is defined as
\begin{equation}\label{eq:imperfect_measurement_record}
    dU = \sqrt{\eta}\,dW + \sqrt{1-\eta}\,dE,
\end{equation}
where $E(t)$ is a Wiener process independent of $W(t)$, and
$\eta \in [0,1]$ denotes the measurement efficiency. 
For $\eta=1$, the observed record coincides with the ideal record, $dU=dW$. 
For $\eta=0$, the observed increment contains no information. 
The Itô products are therefore
\begin{equation*}
    (dU)^2 = dt, \qquad dW\,dU = \sqrt{\eta}\,dt.
\end{equation*}
\begin{figure}[t]
    \centering
    \includegraphics[width=0.48\textwidth]{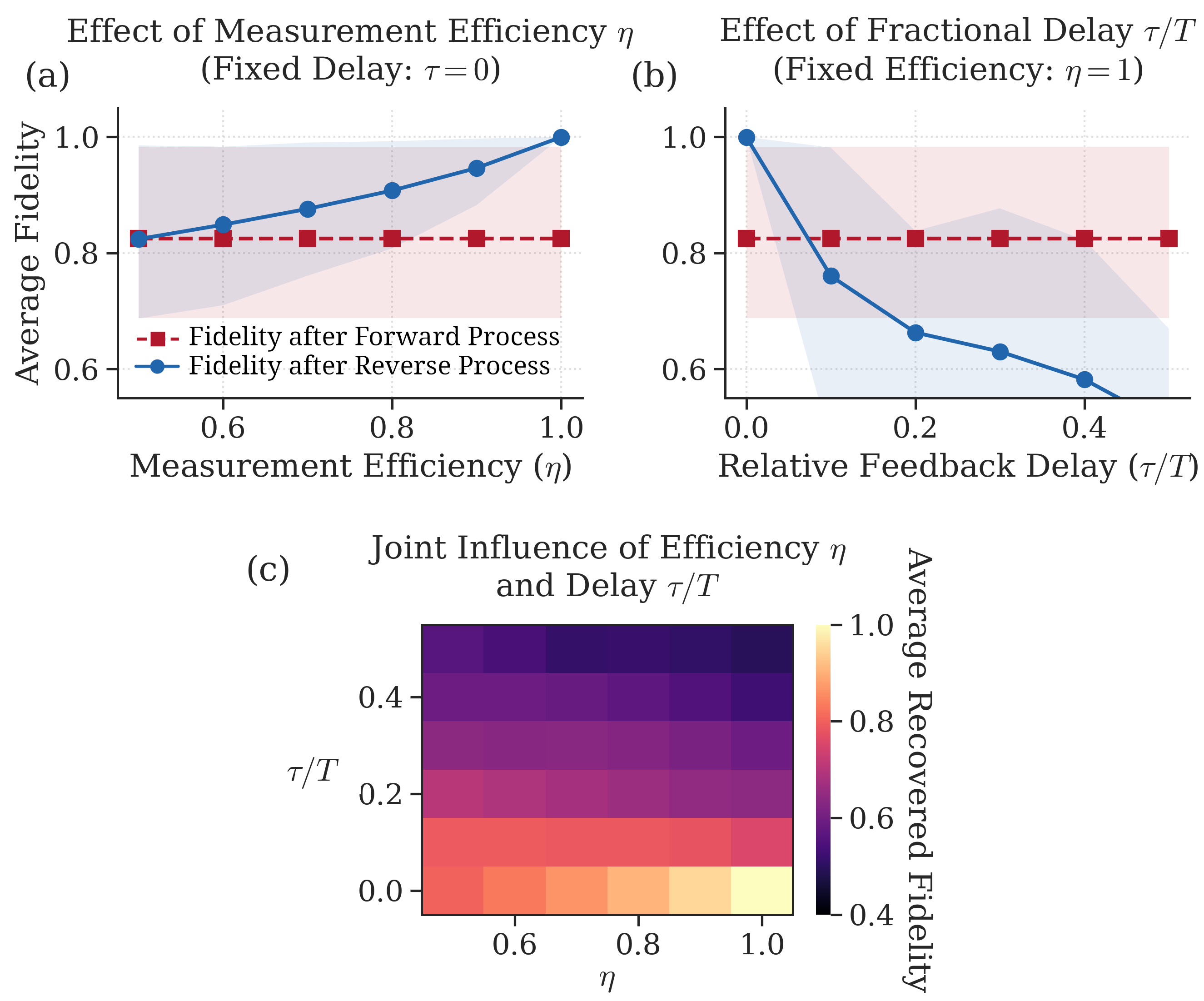}
    \caption{Average state fidelity (blue line) after the reverse process compared with the average fidelity after the forward process (red dashed line). Shaded regions indicate the interval between the 16th and 84th percentiles. 
    (a) Recovered fidelity as a function of measurement efficiency $\eta$ with instantaneous feedback, $\tau=0$. 
    (b) Recovered fidelity as a function of the fractional feedback delay $\tau/T$ with perfect measurement efficiency, $\eta=1$. 
    (c) Recovered fidelity over the joint parameter space $(\eta,\tau/T)$.}
    \label{fig:inef_meas_feedback_delay}
\end{figure}
Thus, for $\eta<1$, the observed process $U(t)$ is only partially correlated with the ideal measurement process $W(t)$. The stochastic evolution generated using $dU$ remains pure at the trajectory level, but it does not coincide pathwise with the actual physical trajectory unless $\eta=1$. The quantitative effect of finite efficiency in continuous-measurement dynamics has been studied in several settings: feedback-based error correction, rapid purification, and monitored quantum metrology \cite{sarovar2004practical,chiruvelli2008rapid,albarelli2018restoring}

We now discuss how feedback delay enters the reverse diffusion equations.
Recall that the reverse SDEs in \cref{eq:simple_rev_sde} generalize the forward SDEs in \cref{eq:simple_fwd_sde} by adding a stochastic feedback term that both accounts for the active noise and drives the state back toward its initial configuration. In the ideal case, this feedback term is evaluated from the instantaneous measurement record. However, with finite delay, the feedback term at time $t$ is instead constructed from the record available at the earlier time $t-\tau$.

We incorporate both feedback delay and measurement inefficiency at the trajectory level as follows. The feedback delay is modelled by evaluating the stochastic driver at the earlier time $t-\tau$, with $0 \leq \tau < T$. Measurement inefficiency is incorporated through the observed increment $dU$ defined in \cref{eq:imperfect_measurement_record}.
Combining these two effects, the reverse SDE becomes
\begin{align}\label{eq:simple_rev_sde_delay_inperf_meas}
    d\ket{\phi(t)} &= \left[\left (-\frac{p}{2}I + A(t)\right)dt + \sqrt{p}\, L \, dW \right] \ket{\phi(t)},  \nonumber \\
    A(t) &= -\frac{X(t-\tau)}{2T-t+\tau}\sqrt{p}L, \nonumber \\
    dX(t) &= -\frac{X(t)}{2T-t}dt + dU, \quad T \leq t \leq 2T .
\end{align}
Because the driver is both delayed and constructed from an imperfect measurement record, the process in \cref{eq:simple_rev_sde_delay_inperf_meas} is no longer an exact reverse process. Exact recovery at $t=2T$ is achieved in the limit $\tau \to 0$ and $\eta \to 1$.

In \Cref{fig:inef_meas_feedback_delay}, we present a qualitative analysis of the initial state recovery under inefficient measurement $\eta$ and feedback delay $\tau$. The forward process decreases the state's fidelity to an average value of $0.8$ (red dashed line). Then, the reverse process in \cref{eq:simple_rev_sde_delay_inperf_meas} is implemented for different values of $\eta$ and $\tau$. For this analysis, we fixed $L = \sigma_1$, $p=0.3$, and $T=1$. From \Cref{fig:inef_meas_feedback_delay}, we can see that the fidelity of the recovered state (blue line) increases with efficiency $\eta$. Conversely, the recovered fidelity decreases as the feedback delay $\tau$ increases, since the feedback term is then constructed from less current information. We note that $\tau/T$ denotes the fractional delay; for example, $\tau/T=0.4$ means that the information entering the feedback term is delayed by $40\%$ of the total process duration.

\section{Discussion}
While the ensemble-average dynamics of open quantum systems are fundamentally irreversible, our results demonstrate that this does not necessarily hold for individual quantum trajectories. The analytical quantum reverse SDEs introduced here establish a theoretical framework that extends classical reverse diffusion theory into the quantum domain. This framework suggests that diffusive effects can not only be reversed but also harnessed as a driver to generate arbitrary diffusion-based quantum gates and, consequently, generate quantum states. This finding opens a pathway for studying quantum generative modelling from first principles of quantum measurement and feedback. Furthermore, the insights from the exact SDEs for simple single-channel Pauli errors provide a foundation for constructing more complex models, such as a reverse SDE for depolarizing noise.

Our work also identifies an interesting open challenge: the robust in situ real-time implementation of the information-dissipative reverse dynamics. While we show that information-conserving dynamics can be reversed in situ via a coherent feedback Hamiltonian, we have thus far only demonstrated [\cref{sec:algo}] a real-time ex situ algorithm for the dissipative case. Achieving a real-time in situ implementation is a critical next step, as it would unlock many interesting applications of quantum diffusion processes.

From the theoretical point of view, it is compelling to relate the reverse SDEs to the continuous-time Petz recovery map \cite{li2025optimality,chen2025recovery,nasu2025quantum,kwon2022reversing} or to the quantum analogue of Bayes' theorem \cite{parzygnat2023time,bai2025quantum,song2025exact,liu2025state,liu2025unifying}, which uses the minimum change principle to determine the forward and reverse processes.

It is also interesting to ask how the present framework can be extended beyond Pauli channels. The derivation of the reverse equations and their corresponding algorithms relied on the special algebraic structure of Pauli operators, which enabled explicit closed-form constructions and algorithmic implementation. For non-Pauli channels, one may be able to derive more general reverse SDEs, but an equally important challenge is to identify whether such equations admit feasible physical or algorithmic realizations. Extending these ideas would therefore require not only new mathematical constructions, but also methods for implementing the resulting reverse dynamics.

\bf Acknowledgements \rm
EG acknowledges support through a grant from the National Research Council of Canada (NRC) and a Canada Graduate Scholarship from the National Science and Engineering Council of Canada (NSERC).

\appendix
\onecolumngrid

\section{The Algorithm for Information-Dissipative Reverse Processes}\label{sec:algo}
\begin{figure}[b]
    \centering \includegraphics[width=0.7\textwidth]{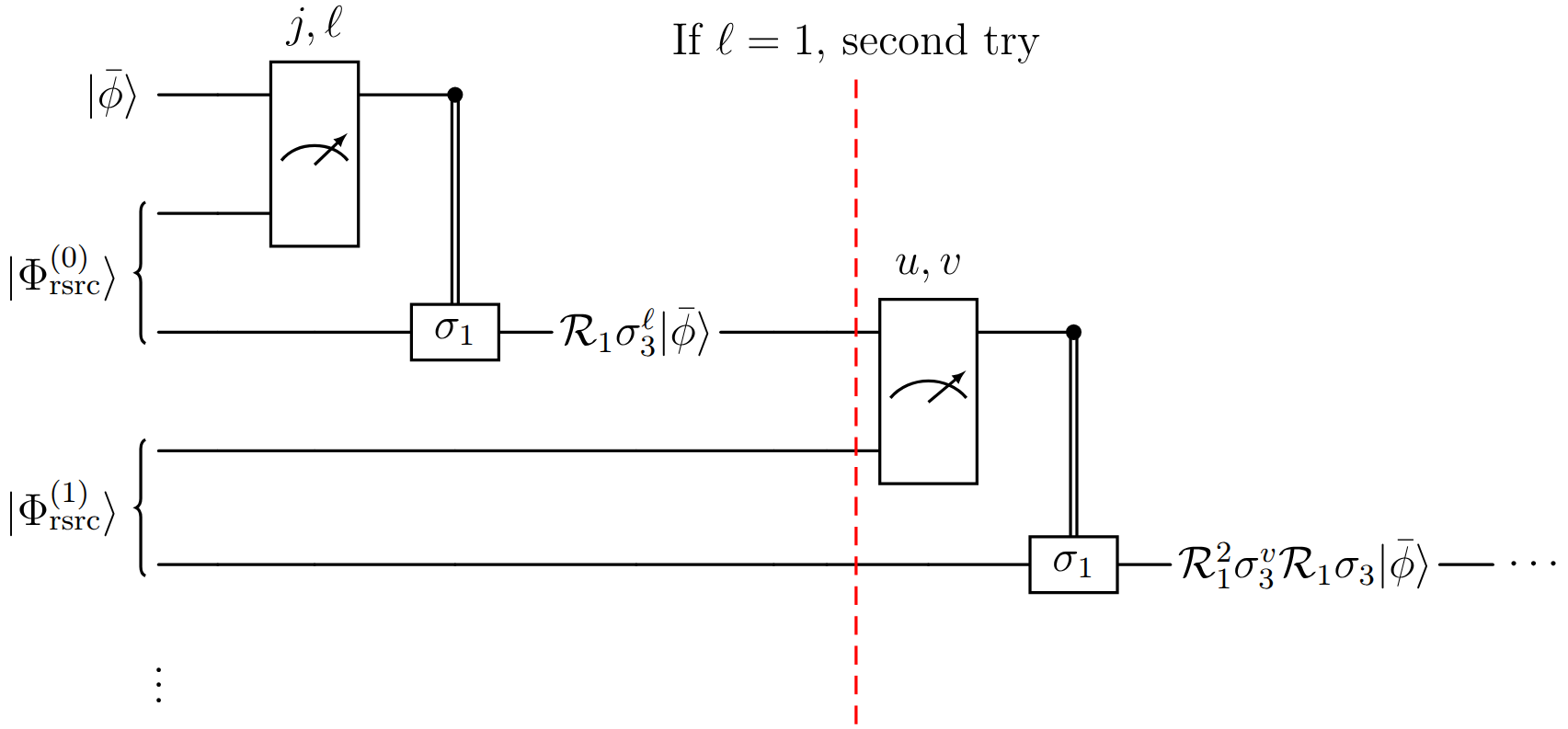}
    \caption{The quantum circuit that near-deterministically implements the stochastic reverse drift $\mathcal{R}_1(t)$. The 2-qubit meter gates implement a Bell-basis measurement with binary outcomes $j$ and $\ell$ ($u$ and $v$). After the measurement, the state of interest acquires the drift $\mathcal{R}_1\sigma_3^{\ell}$, where $\sigma_3^{\ell}$ for $\ell=1$ is an undesirable byproduct. If $\ell=0$, the correct drift is implemented. If $\ell=1$, the process must be repeated.}
    \label{fig:alg_circuit}
\end{figure}
In this section, we introduce an algorithm that implements the stochastic reverse process defined in \cref{eq:ap_simple_rev_sde}. We require the process to be implemented in real time, near-deterministically (controlled error of failure), and without access to the system's state. The main challenge is to implement the stochastic drift, which corresponds to imaginary time evolution (ITE). In what follows, we demonstrate that a series of teleportations and weak measurements generates the entire stochastic reverse process, in real time and near-deterministically, for any given state, without local tomography. We also perform a resource analysis of the algorithm and show that the resource overheads are moderate.

The algorithm can be briefly summarized as follows. For each time interval $\Delta t \ll 1$, we do: Weakly measure the system's state $|\hat \phi(t)\rangle$. This generates a perturbed state $|\bar \phi(t)\rangle$ and the measurement increment $dW$. This is equivalent to realizing one step of the forward process, see \cref{eq:ap_simple_fwd_sde}. Then, the state $|\bar \phi(t) \rangle$ is adjoined with a special two-qubit resource state $|\Phi_{\text{rsrc}}^{(0)}\rangle$ which incorporates $dW$. Then, the system and one resource qubit are measured in the Bell basis. This teleports the system's state and induces the desired ITE stochastic reverse drift $\mathcal{R}_k(t)$. We proceed to the next interval $\Delta t$ and repeat. With the probability slightly more than one-half, the teleportation fails to implement the desired drift due to teleportation byproduct gates. However, this does not mean the state is lost or irreversibly corrupted. In the case of failure, the teleported state is adjoined with another resource state $|\Phi_{\text{rsrc}}^{(1)}\rangle$ and the teleportation is repeated. The repeated teleportation probabilistically cancels the undesirable teleportation byproducts from the previous round and implements the drift. For sufficiently small $\Delta t$, the probability of consecutive failures scales down exponentially. The quantum circuit of the algorithm is shown in \Cref{fig:alg_circuit}.

\textbf{Preliminaries.} The single-qubit forward dynamics are described by the SDE
\begin{align}\label{eq:ap_simple_fwd_sde}
    d\ket{\psi} = \left( -\frac{p}{2} I dt + \sqrt{p} \sigma_k dW \right)\ket{\psi(t)}, \ \ket{\psi(0)}=\ket{\psi_0}.
\end{align}
The corresponding reverse dynamics are governed by
\begin{align}\label{eq:ap_simple_rev_sde}
    d\ket{\phi} &= \left(\left (-\frac{p}{2}I -\frac{X(t)}{2T-t}\sqrt{p} \sigma_k \right)dt + \sqrt{p} \sigma_k dW \right) \ket{\phi(t)}, \ \ \ket{\phi(T)} = \ket{\psi_T}, \nonumber\\
    dX &= -\frac{X(t)}{2T-t}dt + dW, \ \ X(T) = W_T, \ \ T \leq t \leq 2T.
\end{align}
Here, $W_T = W(T)$ denotes the value of the measurement record of the forward process, and $\ket{\psi_T}$ is the corresponding state of unit norm. Equivalently, the reverse evolution can be expressed via the reverse stochastic master equation in \cref{eq:sme_simple_rev} with $L \in \{\sigma_1, \sigma_2, \sigma_3\}$.

\textbf{Setup.} We assume that the forward process in \cref{eq:ap_simple_fwd_sde} extends over the full interval $0 \leq t \leq 2T$.  
Its solution is
\begin{equation}
    F_k(t)\ket{\psi_0}
    = \exp\!\left(-pt + \sqrt{p}\,\sigma_k W(t)\right)\ket{\psi_0}, 
    \ 0 \leq t \leq 2T.
\end{equation}
The solution to the reverse process is given by 
\begin{equation}\label{eq:ap_solution_simple_rev}
    \ket{\phi(t)} = R_k(t)\,F_k(t)\ket{\psi_0}, \ T \le t \le 2T.
\end{equation}
where
\begin{align}\label{eq:ap_reverse_operator_simple}
    R_k(t) :=
    \begin{cases}
        I, \ \ 0 \leq t < T, \\
        \exp\!\Big((X(t)-W(t))\sqrt{p}\,\sigma_k\Big), \ \ T \leq t \leq 2T , \nonumber
    \end{cases}
\end{align}
Furthermore, we have:
\begin{equation}\label{eq:ap_solution_simple_rev_2T}
    \ket{\phi(2T)} \propto \ket{\psi_0}
\end{equation}

\textbf{Goal.} We aim to implement the state $|\hat \phi(t)\rangle = | \phi(t)\rangle / \| | \phi(t)\rangle \|$ for any $t$ in $[T,2T]$. To do so, we partition the interval $[T,2T]$ into subintervals of duration $\Delta t \ll 1$. Within each time step $\Delta t$, we first perform a weak measurement, which generates $\Delta W \approx dW$ and perturbs the state, and then apply the drift, which integrates $\Delta W$ into $\Delta X$. The combined effect is
\begin{equation}
    |\hat \phi(t+\Delta t)\rangle = \frac{\exp ( \Delta Y(t)\sqrt{p}\sigma_k) \exp(-p\Delta t + \sqrt{p}\sigma_k \Delta W) |\hat \phi(t) \rangle}{\| \exp ( \Delta Y(t)\sqrt{p}\sigma_k) \exp(-p\Delta t + \sqrt{p}\sigma_k \Delta W) |\hat \phi(t) \rangle \|},
\end{equation}
where $\Delta Y \approx dX(t) - dW(t)$. This protocol is a discretized version of \cref{eq:ap_solution_simple_rev}, where $\exp(-p\Delta t + \sqrt{p}\sigma_k \Delta W)$ and $\exp(\Delta Y \sqrt{p}\sigma_k)$ correspond to the weak measurement and the imaginary time drift, respectively.

\textbf{Algorithm.} Without loss of generality, we will assume $k=1$ and work in the computational basis. Define a weakly measured state with measurement outcome $\Delta W(t) \approx dW(t)$ as
$$|\bar \phi(t) \rangle_{S_1} = \frac{\exp(-p \Delta t + \sqrt{p}\sigma_k \Delta W(t))| \hat \phi(t)\rangle_{S_1}}{\| \exp(-p \Delta t + \sqrt{p}\sigma_k \Delta W(t))| \hat \phi(t)\rangle_{S_1} \|}.$$
Given the observed increment $\Delta W(t)$ and an unknown state $|\bar \phi(t) \rangle_{S_1}$, we want to apply the following operator
\begin{equation}
    \mathcal{R}_k(t) = \frac{\exp ( \Delta Y(t)\sqrt{p}\sigma_k)}{\| \exp( \Delta Y(t) \sqrt{p}\sigma_k) \| }.
\end{equation}
The key requirement is that the successful application of this operator must be near-deterministic, i.e., the success probability can be made arbitrarily close to $1$ by controlling some parameter. What follows next addresses this challenge. 

To implement $\mathcal{R}_k$ with success probability $1 - \varepsilon$, with $\varepsilon \ll1$, we need to prepare the special resource states that will be consumed upon realization of the operator. For $r=0, 1,\ldots, d-1$, construct a unitary block encoding $U^{(r)}(t)$ that acts on an arbitrary state $\ket{\omega}\ket{0}$ as
\begin{align}
    U^{(r)}(t)(\ket{\omega}\ket{0}) = 
   \mathcal{R}^{2^r}_k(t)\ket{\omega}\ket{0} + \mathcal{B}^{(r)}(t) \ket{\omega}\ket{1},
\end{align}
where $\mathcal{B}^{(r)}(t):= \sqrt{I - (\mathcal{R}^{2^r}_k(t))^2}$. Next, for each $r$ we prepare $N^{(r)}_{\text{Bell}}$ copies of a Bell state, each adjoined with an ancilla state $\ket{0}_A$:
\begin{equation}
    \ket{\Phi}_{B_1B_2}\ket{0}_A := \frac{1}{\sqrt{2}}(\ket{00}_{B_1B_2}+\ket{11}_{B_1B_2})\ket{0}_A
\end{equation}
We simultaneously apply $U^{(r)}(t)$ on all $N^{(r)}_{\text{Bell}}$ copies of $\ket{\Phi}_{B_1B_2}\ket{0}_A$. This yields:
\begin{align}
    \ket{\Phi^{(r)}(t)} &:= (I\otimes U^{(r)}(t))(\ket{\Phi}_{B_1B_2}\ket{0}_A) \nonumber \\
    =\frac{1}{\sqrt{2}}\Big ( &\ket{0}_{B_1} \big ( \mathcal{R}^{2^r}_k(t)\ket{0}_{B_2}\ket{0}_A + \mathcal{B}^{(r)}(t)\ket{0}_{B_2}\ket{1}_A \big )
    + \ket{1}_{B_1} \big ( \mathcal{R}^{2^r}_k(t)\ket{1}_{B_2}\ket{0}_A + \mathcal{B}^{(r)}(t)\ket{1}_{B_2}\ket{1}_A \big ) \Big )
\end{align}
For each $r$, we measure the ancilla $A$ on all $N^{(r)}_{\text{Bell}}$ copies of $\ket{\Phi^{(r)}(t)}$ and post-select only those which had outcome $0_A$. This yields the resource states that ensure the near-deterministic implementation of $\mathcal{R}_k(t)$. The resource states are:
\begin{align}
    \ket{\Phi^{(r)}_{\text{rsrc}}(t)}_{B_1B_2} := \frac{1}{\sqrt{\Pr(0_A)}} \left ( I\otimes \mathcal{R}^{2^r}_k(t) \right )\ket{\Phi}_{B_1B_2}, \text{ for } r=0, 1, \ldots, d-1
\end{align}
It is straightforward to show that $\Pr(0_A) \geq 1/2$. Therefore, to ensure a successful post-selection, for each $r$, we must have $N^{(r)}_{\text{Bell}} \geq \lceil \log_2(1/\delta) \rceil$ where $\delta < 1/2$ is the probability of failing to post-select.

Having prepared the resource states, we are ready to proceed to quantum teleportation. To this end, adjoin $|\Bar \phi(t)\rangle_{S_1}$ with a single copy $\ket{\Phi^{(0)}_{\text{rsrc}}(t)}$. This can be written as:
\begin{align}
    &|\Bar \phi(t)\rangle_{S_1}\ket{\Phi^{(0)}_{\text{rsrc}}(t)}_{B_1B_2}
    = \frac{1}{Z}\sum_{j,\ell=0}^1 \ket{\Phi_{j\ell}}_{S_1 B_1}\mathcal{R}_k(t)  \left ( \sigma_1^j \sigma_3^\ell \right ) |\Bar \phi(t)\rangle_{B_2}
\end{align}
In the above, 
$Z := 2\sqrt{\Pr\bigl(0_A\bigr)}$ and $\ket{\Phi_{j\ell}} := (I \otimes \sigma_1^j\sigma_3^\ell)\,\frac{1}{\sqrt{2}}\,(\ket{00}+\ket{11})$ are Bell basis states for $j,\ell \in \{0,1\}$. We projectively measure qubits $S_1$ and $B_1$ in the Bell basis $\ket{\Phi_{j\ell}}$. If the outcome $(j,\ell)$ is $(0,0)$ or $(1,0)$, we obtain:
\begin{align*}
    \ket{\text{good}^{(0)}} \propto \begin{cases}
        \mathcal{R}_k(t)|\bar \phi(t) \rangle_{B_2}, \text{ for } (j,\ell) = (0,0)\\
        \mathcal{R}_k(t) \sigma_1|\bar \phi(t) \rangle_{B_2}, \text{ for } (j,\ell) = (1,0)
    \end{cases}
\end{align*}
Since we assumed $k=1$, $[\mathcal{R}_k, \sigma_1]=0$ and we can commute $\sigma_1$ to the left. Therefore, both outcomes are equivalent up to a simple Pauli correction by $\sigma_1$. Hence, we successfully implement one step of the reverse process:
\begin{equation}
    \ket{\text{good}^{(0)}} \propto |\hat \phi(t+\Delta t)\rangle_{B_2}
\end{equation}
Now, we can discard all the resource states and proceed to the next increment $\Delta t$, where we perform the weak measurement on the system $B_2$ and implement the ITE drift. Since $B_2$ is the new system's state
we relabel $B_2$ as $S_1$.

Conversely, if the outcome $(j, \ell)$ is $(0,1)$ or $(1,1)$, we get:
\begin{align*}
    \ket{\text{bad}^{(0)}} \propto \begin{cases}
        \mathcal{R}_k(t) \sigma_3|\bar \phi(t) \rangle_{B_2}, \text{ for } (j,\ell) = (0,1)\\
        \mathcal{R}_k(t)  \sigma_1 \sigma_3|\bar \phi(t) \rangle_{B_2}, \text{ for } (j,\ell) = (1,1)
    \end{cases}
\end{align*}
These states are again equivalent up to a Pauli correction by $\sigma_1$. Thus, we have
\begin{equation}
    \ket{\text{bad}^{(0)}} \propto \mathcal{R}_k(t) \sigma_3 |\bar \phi(t) \rangle_{B_2}.
\end{equation}
However, since $[\mathcal{R}_k, \sigma_3] \neq 0$, the outcomes $(0,1)$ and $(1,1)$ do not yield the reverse process state. Indeed, they produce an erroneous state that can not be corrected using unitary operations. For notational convenience, let us relabel $B_2$ as $S_1$ as it is the primary system now. We follow up with another teleportation of $\mathcal{R}^2_k$ by consuming the resource $\ket{\Phi_{\text{rsrc}}^{(1)}}$. Then for $(j,\ell) \in \{(0,1), (1, 1)\}$, we successfully implement the reverse process step:
\begin{align}
    \ket{\text{good}^{(1)}} &\propto \mathcal{R}^2_k(t) \sigma_3 \ket{\text{bad}^{(0)}} \\
    &\propto \mathcal{R}_k(t) \mathcal{R}_k(t) \sigma_3 \mathcal{R}_k(t) \sigma_3|\bar \phi(t) \rangle_{B_2}\\
    &= \lambda_{\text{min}} \mathcal{R}_k(t)|\bar \phi(t) \rangle_{B_2}
\end{align}
In the above, we used the fact that
$$\mathcal{R}_k(t) \sigma_3 \mathcal{R}_k(t) \sigma_3 = \lambda_{\text{min}}I,$$
where $\lambda_{\text{min}}$ is the smallest eigenvalue of $\mathcal{R}_k$.  Also, we ignored the possible teleportation byproduct $\sigma_1$ as it commutes with $\mathcal{R}_k$. For the outcomes $(j,\ell) \in \{(0,0), (1, 0)\}$, we get another erroneous state:
\begin{equation}
    \ket{\text{bad}^{(1)}} \propto \mathcal{R}^2_k(t)\ket{\text{bad}^{(0)}} \propto \mathcal{R}^2_k(t) \mathcal{R}_k(t)\sigma_3 |\bar \phi(t) \rangle_{B_2}
\end{equation}
In the event of failure, we redo the teleportation; however, this time we consume the resource $\ket{\Phi^{(2)}_{\text{rsrc}}(t)}$, which implements $\mathcal{R}^4_k$.  Generally, after $r$ consecutive failures we teleport $\mathcal{R}^{2^r}_k$. The repeated applications $\mathcal{R}^{2^r}_k$  ensure that eventually we obtain the desired state proportional to $\mathcal{R}_k(t)|\hat \phi(t)\rangle$. This is due to the following useful identity. For $k=1$, we have
\begin{equation}\label{eq:ap_teleport_cancel_identity}
    \mathcal{R}^n_k \sigma_3 \mathcal{R}^n_k \sigma_3 = \lambda_{\text{min}}^n I.
\end{equation}

\subsection{Resource Analysis of the Algorithm}\label{sec:ap_resource_analysis}
The proposed algorithm utilizes resource states to ensure that the implementation of the ITE drift is real-time and near-deterministic. While this introduces an overhead, standard methods for implementing ITE are no less resource-intensive. For instance, the proposed ITE algorithms in \cite{motta2020determining,mcardle2019variational,lin2021real,kolotouros2025accelerating,mittal2025deterministic,ray2025quasiprobabilistic,rrapaj2025exact} require partial local tomography of the quantum state at each time step $\Delta t$. This renders the algorithms offline and necessitates substantial offline pre- and post-processing.
Furthermore, approaches such as unitary block encoding with subsequent post-selection are probabilistic; an undesirable measurement outcome irreversibly corrupts the state. In contrast, the proposed algorithm combines unitary block encoding with operator teleportation techniques to realize online, near-deterministic ITE without requiring prior knowledge of the system's state. Below, we evaluate the resource requirements for the proposed algorithm.

At the teleportation attempt $r$ we implement the map $\mathcal{R}_k^{2^r}$. The state-independent upper bound (the worst case) on the probability of teleporting into the wrong branch at the attempt $r$ is
\begin{equation}\label{eq:ap_teleport_fail_probab}
    \Pr(\text{fail}\,|\,r)_{\text{worst}}
    := \frac{1}{2} \left ( 1 + \bigl|\tanh \bigl(2^{r+1}\sqrt{p} \, \Delta Y \bigr )\bigr | \right)
    = \frac{1}{2}\Bigl( 1 + 2^{r+1} \sqrt{p}\, |\Delta Y| \Bigr )
      + O \bigl( 2^{3(r+1)} p^{3/2} |\Delta Y|^3 \bigr ).
\end{equation}
Note that $\Delta Y$ is of order $\Delta t$, which implies $|\Delta Y|^3$ is of order $\Delta t^3$.

To see that \cref{eq:ap_teleport_fail_probab} is true, we note that for each fixed attempt $r$ the good and bad teleportation outcomes can be viewed as unnormalized POVMs:
\begin{equation}
    E_{\text{good}}^{(r,\text{unnorm})}
    = \left ( K_1^{(r)} \right )^{\dagger}K_1^{(r)} + \left ( K_2^{(r)} \right )^{\dagger}K_2^{(r)},
    \qquad
    E_{\text{bad}}^{(r,\text{unnorm})}
    = \left ( K_3^{(r)} \right)^{\dagger}K_3^{(r)} + \left (K_4^{(r)} \right)^{\dagger}K_4^{(r)},
\end{equation}
where the Kraus operators at the attempt $r$ are
\begin{equation}
    K_1^{(r)} := \mathcal{R}_k^{2^r}, \quad
    K_2^{(r)} := \mathcal{R}_k^{2^r}\sigma_1, \quad
    K_3^{(r)} := \mathcal{R}_k^{2^r}\sigma_1\sigma_3, \quad
    K_4^{(r)} := \mathcal{R}_k^{2^r}\sigma_3.
\end{equation}
Since these POVMs are unnormalized, we obtain
\begin{equation}
    E_{\text{good}}^{(r,\text{unnorm})}
    + E_{\text{bad}}^{(r,\text{unnorm})}
    = 4e^{-2^{r+1}\sqrt{p}\,|\Delta Y|}\,
      \cosh \bigl( 2^{r+1}\sqrt{p}\,\Delta Y(t) \bigr)\,I.
\end{equation}
Normalizing by $4e^{-2^{r+1}\sqrt{p}\,|\Delta Y|}\cosh\bigl(2^{r+1}\sqrt{p}\,\Delta Y(t)\bigr)$ yields the bad branch POVM at the attempt $r$:
\begin{equation}
    E_{\text{bad}}^{(r)}
    = \frac{1}{2} \Bigl( I - \tanh \bigl( 2^{r+1}\sqrt{p}\,\Delta Y(t) \bigr)\sigma_k \Bigl).
\end{equation}
The eigenvalues of $E_{\text{bad}}^{(r)}$ are
\begin{equation}
    \lambda_{\pm}\bigl(E_{\text{bad}}^{(r)}\bigr)
    = \frac{1}{2} \Bigl( 1 \pm \tanh \bigl( 2^{r+1}\sqrt{p}\,\Delta Y(t) \bigr) \Bigl).
\end{equation}
For the worst case we take the largest eigenvalue,
\begin{equation}
    \Pr(\text{fail}\,|\,r)_{\text{worst}}
    = \lambda_{+}\left ( E_{\text{bad}}^{(r)} \right ),
\end{equation}
which gives \cref{eq:ap_teleport_fail_probab}.

Having computed the probability of a single teleportation failure at the attempt $r$, we now consider the probability of reaching the $r$-th consecutive failure (i.e., failing at all attempts $0,1,\ldots,r-1$). This probability is bounded as
\begin{equation}
    \Pr(r \ \text{consecutive fails})_{\text{worst}}
    \leq \prod_{s=0}^{r-1} \Pr(\text{fail}\,|\,s)_{\text{worst}}
    = \prod_{s=0}^{r-1} \frac{1}{2}
      \left(1+\bigl|\tanh \bigl(2^{s+1}\sqrt{p}\,\Delta Y\bigr)\bigr|\right).
\end{equation}
For small $|\Delta Y|$ the first-order estimate is
\begin{equation}
    \Pr(r \ \text{consecutive fails})_{\text{worst}}
    \lesssim \left (\frac{1}{2} \right)^r \,
      \prod_{s=0}^{r-1} \Bigl(1+2^{s+1}\sqrt{p}\,|\Delta Y(t)|\Bigr).
\end{equation}
Recall that we allow at most $d$ teleportation attempts, $r=0,1,\ldots,d-1$. To ensure that the total failure probability is at most $\varepsilon$, it suffices to choose $d$ such that
\begin{equation}
    \Pr(d \ \text{consecutive fails})_{\text{worst}} \leq \varepsilon.
\end{equation}
A convenient sufficient condition is obtained by bounding all factors by their largest value, which occurs at the attempt $d-1$. Define
\begin{equation}
    \eta := \left |\tanh\bigl(2^{d}\sqrt{p}\,\Delta Y(t)\bigr) \right |.
\end{equation}
Then
\begin{equation}
    \Pr(d \ \text{consecutive fails})_{\text{worst}}
    \leq  \left ( \frac{1}{2} \right )^d \bigl(1+\eta \bigr)^d .
\end{equation}
Consequently, a sufficient choice of $d$ is any integer satisfying
\begin{equation}
     \left ( \frac{1}{2} \right )^d \bigl(1+\eta \bigr)^d \leq \varepsilon.
\end{equation}
Equivalently, we can define the worst-case minimal budget of post-selected resource states implicitly as
\begin{equation}
    d_{\min}
    := \min \left \{ d \in \mathbb{N} \, \middle | \,
    d \, \bigl ( 1 - \log_2( 1 + \eta ) \bigr )
    \geq \log_2 \left( \frac{1}{\varepsilon} \right) \right \}.
\end{equation}
For $2^{d}\sqrt{p}\,|\Delta Y(t)| \ll 1$ we can use
$\log_2(1 + \eta) \approx \log_2(e) \, \eta
\approx \log_2(e)\, 2^{d}\sqrt{p} \, |\Delta Y(t)|$,
which yields the implicit approximate equation
\begin{equation}
    d_{\min} \approx
    \left \lceil
        \frac{\log_2 \left(\frac{1}{\varepsilon} \right)}
             {1 - \log_2(e)\,2^{d}\sqrt{p}\,|\Delta Y(t)|}
    \right \rceil.
\end{equation}
In the limit $\Delta t \rightarrow 0$ (so that $|\Delta Y(t)|\to 0$), we recover
$d_{\min} \rightarrow \lceil \log_2(1/\varepsilon) \rceil$.

\subsection{Generalization to Multi-Qubit Pauli Errors.}
It is straightforward to generalize the algorithm to multiqubit $m$-local Pauli error, i.e., $L\in \{I, \sigma_1, \sigma_2, \sigma_3\}^{\otimes m}$. In this case, 
we define
\begin{align}\label{eq:ap_reverse_operator_multi_qubit_simple}
    R(t) :=
    \begin{cases}
        I, \ \ 0 \leq t < T, \\
        \exp \left(\Delta Y(t)\sqrt{p} L \right ) \ \ T \leq t \leq 2T , \nonumber
    \end{cases}
\end{align}
and $\mathcal{R}(t) := R(t)/||R(t)||$. The resource state is
\begin{equation}
    \ket{\Phi^{(r)}(t)} = (I \otimes U^{(r)}(t))(\ket{\Phi}^{\otimes m}\ket{0}),
\end{equation}
where $U^{(r)}(t)$ is a block encoding of $\mathcal{R}^{2^r}(t)$, and $\ket{\Phi}$ denotes the Bell state. The teleportation byproduct is a Pauli string given by
\begin{equation}\label{eq:ap_multiqubit_byproduct}
    P:=\bigotimes_{u=1}^m \sigma_1^{j_u}\sigma_3^{\ell_u},
\end{equation}
with $j,\ell \in \{0,1\}^m$. The teleportation byproducts can be divided into two branches. The good branch -- byproducts that commute with  $\mathcal{R}$ and hence can be corrected, and the bad branch -- byproducts that do not commute with $\mathcal{R}$ and hence require repeated teleportation.

As before, to cancel undesirable byproducts, we repeat the teleportation. The following identity is the generalization of \cref{eq:ap_teleport_cancel_identity}, and it can be used to annihilate teleportation byproducts.
\begin{equation}\label{eq:ap_general_teleportation_byprod}
    \mathcal{R}^n P_* \mathcal{R}^n P_* = \lambda^n I,
\end{equation}
where $P_*$ is a fixed Pauli string which does not commute with $\mathcal{R}$ and $\lambda$ is the lowest eigenvalue of $\mathcal{R}$. Then, for any teleportation byproduct $P$ that does not commute with $L$ (and hence with $\mathcal{R}$), there is the Pauli string $C= P_*P$ that commutes with $L$  such that
\begin{equation}
    P = P_*C.
\end{equation}
Therefore, if the teleportation yields  $\mathcal{R}P$, then we can write
\begin{equation}
    \mathcal{R}P = \mathcal{R}P_*C = s \mathcal{R}CP_* = s C\mathcal{R}P_*,
\end{equation}
where $s \in \{+1, -1\}$ is the result of commuting $P_*$ and $C$. Since $C$ commutes with $\mathcal{R}$, it is straightforward to correct it, and the global phase $s$ can be dropped. It follows that up to the correction by $C$, we get the following equivalence:
\begin{equation}
    \mathcal{R}P \equiv \mathcal{R}P_*.
\end{equation}
This equivalence shows that undesirable teleportation result $\mathcal{R}P$ with $P$ not commuting with $\mathcal{R}$ can be reduced to $\mathcal{R}P_*$, and subsequently further treated through the use of repeated teleportation that takes advantage of the identity in \cref{eq:ap_general_teleportation_byprod}.

We now argue that the probability of obtaining a bad teleportation byproduct does not depend on the locality $m$ of $L$. 
Since $L^2 = I$, we can write
\begin{equation}
    \mathcal{R}(t) = a(t) I + b(t) L,
\end{equation}
for some scalars $a(t), b(t)$ determined by $\sqrt{p}\Delta Y(t)$. For any Pauli byproduct $P$, the corresponding Kraus operator is $K_P \propto \mathcal{R}(t)P$, and a direct calculation shows that
\begin{equation}
    K_P^\dagger K_P = \alpha_\pm I + \beta_\pm L,
\end{equation}
where the choice of sign $\pm$ (and hence the coefficients $\alpha_\pm, \beta_\pm$) depends only on whether $P$ commutes or anticommutes with $L$. Therefore, the POVM elements associated with the good and bad branches,
\begin{equation}
    E_{\text{good}} = \sum_{P:\,[P,L]=0} K_P^\dagger K_P,
    \qquad
    E_{\text{bad}} = \sum_{P:\,\{P,L\}=0} K_P^\dagger K_P,
\end{equation}
are also of the form $\alpha I + \beta L$. Their eigenvalues depend only on the eigenvalues $\pm 1$ of $L$, and thus are independent of the Hilbert space dimension and of $m$. Consequently, the worst-case failure probability $\Pr(\text{fail}\,|\,r)_{\text{worst}}$ is the same function of $\sqrt{p}\Delta Y$ as in the single-qubit case,
\begin{equation}
\Pr(\text{fail}\,|\,r)_{\text{worst}}
    = \frac{1}{2} \left ( 1 + \bigl|\tanh \bigl(2^{r+1}\sqrt{p} \, \Delta Y \bigr )\bigr | \right)
    \approx \frac{1}{2}\Bigl( 1 + 2^{r+1} \sqrt{p}\, |\Delta Y| \Bigr )
\end{equation}
and does not depend on the number of qubits in the support of $L$.

\section{Constructing the Reverse Depolarizing Noise SDE}\label{sec:app_rev_dn}
In this section, we construct the reverse SDE for depolarizing noise. We begin with the forward, information-dissipative It\^o SDE:
\begin{align}\label{eq:ap_forward_dn_dissipative} d\ket{\psi(t)} &= \left(-\frac{p}{2} I dt + \sum_{k=1}^3\sqrt{\frac{p}{3}}\sigma_k dW_k(t) \right) \ket{\psi(t)}, \ 0 \leq t \leq T, \nonumber \\
\ket{\psi(0)} &= \ket{\psi_0}
\end{align}
Using the second-order stochastic Magnus expansion for linear Itô SDEs with constant coefficients, we obtain an approximate solution:
\begin{equation}\label{eq:mag2_state}
    |\psi^{(2)}(t)\rangle := \exp(-p I t + \sqrt{\frac{p}{3}} \sum_{k=1}^3 \sigma_k W_k(t) + i \frac{2p}{3}( \sigma_1 S_{23}(t) + \sigma_2 S_{31}(t) + \sigma_3 S_{12}(t)))\ket{\psi_0}
\end{equation}
Here, $S_{ij}(t)$ are Lev\'y's stochastic areas defined as
\begin{equation}
    S_{ij}(t) = \frac{1}{2}\int_0^t (W_i(s)dW_j(s) - W_j(s)dW_i(s) ).
\end{equation}
The second-order truncation error is controlled in the $L_2$ sense as
\begin{equation}\label{eq:L2_error}
    \left( \mathbb{E} \left [ \left \| \ket{\psi(t)} - |\psi^{(2)}(t)\rangle \right \|_2^{2} \right ] \right)^{1/2}
    = O \left( p^{3/2} t^{3/2} \right),
\end{equation}
see, e.g., \cite{wang2020magnus}. 
Let $F^{(2)}(t)$ denote the second-order Magnus expansion operator,
\begin{equation}
    F^{(2)}(t) := \exp(-p I t + \sqrt{\frac{p}{3}} \sum_{k=1}^3 \sigma_k W_k(t) + i \frac{2p}{3}( \sigma_1 S_{23}(t) + \sigma_2 S_{31}(t) + \sigma_3 S_{12}(t))).
\end{equation}
Define the operator $R^{(2)}(t)$ as
\begin{equation}\label{eq:ap_rev_apporx_op}
    R^{(2)}(t) := 
    \begin{cases}      
        I, \quad  0 \leq t < T, \\
        \exp(-p I (t-T) + \sum_{k=1}^3 \sigma_k (X_k(t) - X_k(T))), \ \  T \leq t \leq 2T,
    \end{cases}
\end{equation}
where, for $k=1,2,3$, the complex stochastic processes $X_k(t)$ and their boundary conditions $X_k(T)$ and $X_k(2T)$ are defined as
\begin{align}\label{eq:ap_boundary_conditions}
    X_k(t) = \sqrt{\frac{p}{3}} U_k(t) + i \frac{2p}{3} V_k(t), \quad
    X_k(T) = \sqrt{\frac{p}{3}} W_k(T) + i \frac{2p}{3} S_{g(k)}(T), \quad
    X_k(2T) = 0.
\end{align}
Here, the function $g(k)$ provides the index mapping for the Lévy areas: $g(1)=(2,3)$, $g(2)=(3,1)$, and $g(3)=(1,2)$. The processes $U_k(t)$ and $V_k(t)$ are stochastic processes that satisfy the boundary conditions above and
\begin{equation}\label{eq:ap_condition_3}
    dU_k^2 = dV_k^2 = dt, \ \ dU_i dV_j = \delta_{ij}dt.
\end{equation}
Then it follows that
\begin{equation}
    R^{(2)}(2T)|\psi(T)\rangle = R^{(2)}(2T)F^{(2)}(T)\ket{\psi_0} + O \left( p^{3/2} T^{3/2} \right) = e^{-p2T}\ket{\psi_0} + O \left( p^{3/2} T^{3/2} \right).
\end{equation}
Note that $\ket{\psi(t)}$ is the exact solution to the forward process, and we have $\ket{\psi(t)} = F^{(2)}(t)\ket{\psi_0}+ O \left( p^{3/2} t^{3/2} \right)$. Furthermore, $R^{(2)}(2T)F^{(2)}(T) = e^{-p2T}I$ due to \cref{eq:ap_boundary_conditions}.

Finally, we use It\^o's lemma to compute the reverse SDE. To this end, we need to calculate the first and second derivatives of $R^{(2)}(t)$. The derivative of the matrix exponential $e^{\chi(t)}$ is
\begin{equation}
    \frac{\partial}{\partial x}e^{\chi(t)} = \frac{e^{\text{ad}_{\chi}} -1}{\text{ad}_{\chi}} \left ( \frac{\partial}{\partial x} \chi(t) \right) e^{\chi(t)},
\end{equation}
where $\text{ad}_{\chi}(\cdot) := [\chi, \cdot]$, see \cite{blanes2009magnus}. Note that the derivatives are given by the Taylor series of the exponential of $\text{ad}_{\chi}(\cdot)$. To keep the results analytically tractable, we  truncate the series as:
\begin{align}\label{eq:ap_truncated_derivative}
    \frac{e^{\text{ad}_{\chi}} -1}{\text{ad}_{\chi}} \left ( \frac{\partial}{\partial x} \chi(t) \right) e^{\chi(t)} = \sum_{n=0}^{\infty}\frac{\mathrm{ad}_{\chi}^{n}}{(n+1)!} \left ( \frac{\partial}{\partial x} \chi(t) \right) e^{\chi(t)} 
    \approx \left (I + \frac{1}{2}\text{ad}_{\chi} \right ) \left ( \frac{\partial}{\partial x} \chi(t) \right) e^{\chi(t)}
\end{align}
Let the exponent of $R^{(2)}$ be denotes as
\begin{equation}
    \chi(t) = -p I (t-T) + \sum_{k=1}^3 \sigma_k (X_k(t) - X_k(T)).
\end{equation}
Then, applying It\^o's lemma to $R^{(2)}(t)$ with the truncated derivatives [\cref{eq:ap_truncated_derivative}] yields the following reverse SDE:
\begin{align}\label{eq:ap_dn_general_rev_sde}
    d\ket{\phi(t)} = -p \ket{\phi(t)} dt &+ \sum_{k=1}^3 \left ( \sigma_k
    + \frac{1}{2}\sum_{j=1}^3 [\sigma_j, \sigma_k](X_j(t)-X_j(T)) \right) \ket{\phi(t)} dX_k \nonumber \\
    &+ \frac{1}{2} \sum_{k=1}^3 \left ( 1 - \sum_{j\neq k}^3 (X_j(t) - X_j(T))^2 \right )\ket{\phi(t)} dX_k^2, \ \ T \leq t \leq 2T
\end{align}
Under the boundary conditions in \cref{eq:ap_boundary_conditions}, a Brownian bridge provides the simplest model for the dynamics of the stochastic variables $U_k(t)$ and $V_k(t)$. Therefore, we assume that the dynamics of $U_k(t)$ and $V_k(t)$ are governed by Brownian bridge SDEs with the shared stochastic measurement increment $dW_k(t)$. This is a reasonable but simplifying assumption that allows us to reduce the stochastic dynamics to three Brownian bridge-like processes:
\begin{align}
    dX_k(t) &= \sqrt{\frac{p}{3}}d U_k(t) + 2i \frac{p}{3}dV_k(t) \nonumber \\
    &= \sqrt{\frac{p}{3}} \left (-\frac{U_k(t)}{2T-t}dt + dW_k(t) \right ) + 2i \frac{p}{3} \left ( -\frac{V_k(t)}{2T-t}dt + dW_k(t) \right ) \nonumber \\
    &= -\frac{X_k(t)}{2T-t}dt + \left( \sqrt{\frac{p}{3}} + 2i \frac{p}{3} \right ) dW_k(t)
\end{align}
Hence, $X_k(t)$ for $k=1,2,3$ is a Brownian bridge with the following convenient properties:
\begin{equation}
    dX_i dX_j(t) = \delta_{ij}\left ( \sqrt{\frac{p}{3}} + 2i\frac{p}{3} \right)^2 dt.
\end{equation}
Therefore, for $T \leq t \leq 2T$, the final reverse SDE is
\begin{align}\label{eq:ap_dn_final_rev_sde}
    d\ket{\phi(t)} = -p \ket{\phi(t)} dt &+ \sum_{k=1}^3 \left ( \sigma_k + \frac{1}{2}\sum_{j=1}^3 [\sigma_j, \sigma_k](X_j(t)-X_j(T)) \right) \ket{\phi(t)} dX_k \nonumber \\
    &+ \frac{\gamma^2}{2} \sum_{k=1}^3 \left ( 1 - \sum_{j\neq k}^3 (X_j(t) - X_j(T))^2 \right ) \ket{\phi(t)} dt,
\end{align}
where for $k=1,2,3$, we have
\begin{align}
    dX_k &= -\frac{X_k(t)}{2T-t}dt + \gamma dW_k(t), \ \  X_k(T) = \sqrt{\frac{p}{3}} W_k(T) + 2i \frac{p}{3} S_{g(k)} (T), \\ 
    \gamma &= \sqrt{\frac{p}{3}} + 2i\frac{p}{3}.
\end{align}
Let us now examine the error caused by truncating the derivative in \cref{eq:ap_truncated_derivative}. The series that we have truncated are
\begin{equation}
    \mathcal{E}(t)=\sum_{n\ge2}\frac{\mathrm{ad}_{\chi(t)}^{\,n}}{(n+1)!} = \frac{e^{\text{ad}_{\chi(t)}}-1}{\text{ad}_{\chi(t)}} - 1 - \frac{\text{ad}_{\chi(t)}}{2}.
\end{equation}
Using the identity
\begin{equation}
    \frac{e^x-1}{x} - 1 - \frac{x}{2} \leq \frac{x^2}{3!} \frac{e^x-1}{x}, \quad x > 0,
\end{equation}
we can deduce that
\begin{equation}
\|\mathcal{E}(t)\| \leq \frac{2}{3!}\| \tilde \chi(t) \| \left ( e^{2\| \tilde \chi(t) \|} - 1\right),
\end{equation}
where $\tilde \chi(t) := \sum_{k=1}^3 \sigma_k (X_k(t) - X_k(T))$. In the regime $\|\tilde \chi(t)\| < pT < 1$, the bound simplifies to
\begin{align}
    \|\mathcal{E}(t)\| &\leq \frac{2}{3!}\| \tilde \chi(t) \| \left ( 1 + 2\|\tilde \chi(t) \| - 1 + O(\| \tilde \chi(t) \|^2)\right) \\
    &= \frac{1}{3}\| \tilde \chi(t) \|^2 + O(\| \tilde \chi(t) \|^3) = O(\| \tilde \chi(t) \|^2).
\end{align}
Then, for some positive constant $c$, the total error accumulates on the time interval $[T,2T]$ as
\begin{equation}
    E_{\text{total}} \leq \left ( c \ \mathbb{E} \left | \int_T^{2T} \| \tilde \chi(t) \|^2 dX_k \right |^2 \right)^{\frac{1}{2}} = O\left(  p^{\frac{3}{2}} T^{\frac{3}{2}} \right).
\end{equation}
The truncated second derivatives contribute a negligible amount of accumulated error, which is the square of the error above.

To obtain the final total error, we must consider two stages of the construction of the reverse SDE. First, the operator $R^{(2)}(t)$ is the reverse of the forward operator $F^{(2)}(t)$, which is the second-order Magnus approximation of the forward process. Essentially, $R^{(2)}(t)$ reverses a slightly different forward process which deviates from the original forward process by a root mean squared error $O(p^\frac{3}{2} T^\frac{3}{2})$. Second, to derive the reverse SDE, we apply It\^o's lemma to $R^{(2)}(t)$. To keep calculations tractable, we use truncated derivatives, which yield an approximate reverse SDE with the root mean squared error $O(p^\frac{3}{2} T^\frac{3}{2})$. Therefore, for $pT \leq 1$, the total root mean squared error of the reverse process grows as $O(p^\frac{3}{2} T^\frac{3}{2})$. It follows that the reverse process can recover the initial state $\ket{\psi_0}$ of the forward process with the mean squared error
\begin{equation}
    \mathbb{E}[ \| \, |  \hat \phi(2T) \rangle - \ket{\psi_0}\|^2 ] = O \left ( p^3 T^3 \right ),
\end{equation}
where $|\hat \phi(2T)\rangle$ denotes the normalized state of $\ket{\phi(2T)}$.
Since for any unit vectors $\ket{u}$ and $\ket{v}$,
\begin{equation*}
    1-\left|\langle u|v\rangle\right|
    \leq
    \frac{1}{2}\|\,|u\rangle-|v\rangle\,\|^2,
\end{equation*}
the expected fidelity between  $|\hat \phi(2T)\rangle$ and $\ket{\psi_0}$ is given by
\begin{align}
   \mathbb{E}\left [ F \left (|\hat \phi(2T)\rangle, \ket{\psi_0} \right ) \right ] &:= \mathbb{E} \left [\left | \langle \hat \phi(2T)| \psi_0 \rangle \right | \right ] \geq 1 - O(p^3T^3).
\end{align}

\bibliography{refs}

\end{document}